\documentclass[reprint, superscriptaddress, amsmath,amssymb, aps, pre, floatfix]{revtex4-1}

\usepackage{graphicx} % Include figure files
\usepackage{xcolor}   % Add coloring options
\usepackage{hyperref} % add hypertext capabilities
\hypersetup{          % set up additional options for hyperref
 pdftitle={Numerical Study of Continuous and Discontinuous Dynamical Phase Transitions for Boundary Driven Systems},
 pdfauthor={Ohad Shpielberg, Yaroslav Don, and Eric Akkermans},
 pdfsubject={},
 pdfkeywords={},
 unicode=true,
 breaklinks=true,
 linkcolor=blue, 
 citecolor=[rgb]{0.0,0.7,0.1},
 pdfborder={0 0 0},
 pdfborderstyle={},
 colorlinks=true,
 bookmarksnumbered=true,
 bookmarksopen=true,
 bookmarksopenlevel=1,
}

\usepackage[all]{xy}                        % typesetting graphs & diagrams
\SelectTips{cm}{10}                         % better tips
\newcommand*{\particle} {\displaystyle\bigcirc}       % particle (blank)
\newcommand*{\vacancy}  {\phantom{\particle}}         % no particle
\newcommand*{\fullsite} {\dfrac{\particle}{}}         % site w/  particle
\newcommand*{\emptysite}{\frac{\vacancy}{}}           % site w/o particle

\DeclareMathOperator{\E}{e}                           % Euler's number (e)
\newcommand*{\D}{\operatorname{d}\!}                  % differential   (dx)
\newcommand*{\I}{\mkern1.5mu\mathrm{i}\mkern1.5mu}    % imaginary unit (i)

\newcommand*{\mean}[1]{\left\langle #1 \right\rangle} % Operator value <O>

\newcommand*{\eg}{e.g., }                   % proper punctuation of "e.g."
\newcommand*{\aesthetic}{\ae{}sthetic }     % use Latin æsc for "aesthetic"

\begin{document}

\title{Numerical Study of Continuous and Discontinuous Dynamical Phase Transitions for Boundary Driven Systems}

\author{Ohad Shpielberg}
\email{ohad.shpielberg@lpt.ens.fr}
\affiliation{Laboratoire de Physique Th\'{e}orique, \'{E}cole Normale Sup\'{e}rieure and CNRS, 75005 Paris, France}
\affiliation{Physics Department, Technion--Israel Institute of Technology, 3200003 Haifa, Israel}

\author{Yaroslav Don}
\affiliation{Physics Department, Technion--Israel Institute of Technology, 3200003 Haifa, Israel}

\author{Eric Akkermans}
\affiliation{Physics Department, Technion--Israel Institute of Technology, 3200003 Haifa, Israel}

\date{\today}

% ======================================================================================== %

\begin{abstract}
The existence and search for thermodynamic phase transitions is of unfading interest. In this paper, we present  numerical evidence of dynamical phase transitions occurring in boundary driven systems with a constrained integrated current. It is shown that certain models  exhibit a discontinuous transition between two different density profiles and a continuous transition between a time-independent and a time-dependent profile. We also checked that the KMP model does not exhibit phase transition in a range much larger than previously explored.
\end{abstract}

\maketitle

% ======================================================================================== %

\section{Introduction
\label{sec:introduction}}

Out-of-equilibrium systems are currently the object of considerable attention both in classical and in quantum physics~\cite{Mallick2015,Bernard2016,Pilgram2003,Karrasch2013,Mendoza2013,Prosen2009,Zotos1997,Saito2011a}. An important aspect of  out-of-equilibrium physics resides in the study of non-equilibrium steady states of boundary driven systems (\eg two reservoirs at non-equal densities), and corresponding fluctuations about the steady state.  A useful approach in the study of non-equilibrium steady states is a hydrodynamic description known as the Macroscopic Fluctuation Theory (MFT)~\cite{Bertini2015,Bertini2009,Bertini2003}.  The MFT  provides an efficient way to obtain an expression for the probability of observing steady state fluctuations. As such, the MFT has proven useful in obtaining various properties such as out-of-equilibrium density fluctuations~\cite{Bertini2007,Aminov2014, Baek2015}, Clausius inequality~\cite{Bertini2013}, emergence of void formations~\cite{Krapivsky2012a}, and others~\cite{Akkermans2013,Bodineau2010a,Bodineau2008a,Bodineau2010,Agranov2016,Krapivsky2014a}. 

Another important problem in that field is the characterization of current fluctuations. Current fluctuations have been thoroughly studied in statistical physics~\cite{Derrida2004,Gorissen2012,Appert-Rolland2008,Tizon16a,Tizon16b}  as well as in mesoscopic physics~\cite{Akkermans2007,Kamenev2011,Jordan2004}. Knowledge of current fluctuations allows to measure how close is a fluctuation to the steady state. Moreover, the noisy nature of the measurement allows to obtain information about the system under study~\cite{Kambly2011}. 

Current fluctuations can be obtained from the probability $P_t\left(Q\right)$ that a net amount of $Q$ particles (or any other amount, \eg heat) flowed through the system during a time $t$. Generally, $P_t\left(Q\right)$  is dominated by a single fluctuation. However, finding it is a hard optimization problem~\cite{Bertini2006}. It was conjectured in~\cite{Bodineau2004} that the dominant trajectory  is time-independent. This is the content of the Additivity Principle (AP), which allows to simplify the aforementioned minimization problem, and has proven useful~\cite{Bertini2006,Imparato2009,Saito2011a,Hurtado2009,Hurtado2010,Gorissen2012}.  

In certain cases, the AP solution may become non-unique, or overtaken by a time-dependent solution. In boundary driven systems, only until recently~\cite{Baek2016b} there were no known physical examples of a continuous transition between AP solutions.  Moreover, as of yet there is no example of a continuous transition from an AP solution to a time-dependent one (namely, breaking of the AP assumption~\footnote{However, for periodic boundary conditions, many models were shown to break the AP assumption explicitly, see~\cite{Appert-Rolland2008} and~\cite{Zarfaty2015a,Bodineau2005,Hurtado2011a}. }),
as well as no example of a discontinuous transition between two distinct AP solutions.  

By analogy with thermodynamic phase transitions, it is useful to interpret a breaking of the AP assumption or a discontinuous transition as a dynamical phase transition (DPT) of first and second order, respectively, due to the non-analyticity of $P_t\left(Q\right)$  at the transition point~\cite{Bertini2005,Imparato2009,Shpielberg2016}. 
This interpretation has been used in~\cite{Shpielberg2016} to help obtaining a sufficient and necessary condition for the validity of the AP for small fluctuations. Note that while one dimensional equilibrium thermodynamic systems with short-range interactions never display phase transitions~\cite{LandauLifshiftz1980StatisticalPt1}, constrained systems---in or out of equilibrium---may very well present them~\cite{Appert-Rolland2008,Zarfaty2015a,Bodineau2005}.

The purpose of this paper is to numerically implement the tools developed in~\cite{Shpielberg2016} for some models of interest. 
First, we report a thorough numerical study of the Kipnis--Marchioro--Presutti (KMP) model~\cite{Kipnis1982}, which shows that it never violates the sufficient and necessary conditions given in~\cite{Shpielberg2016} and for a much broader parameter range than previously explored. 
Second, we present numerical evidence for either first or second order DPT as a function of the strength of a constrained current in some boundary driven systems. 
A physical example for such a process is the Long-Range Hopping with Exclusion model proposed by Bodineau~\cite{BodineauPrivate}. 

The outline of this paper is as follows. In Sec.~\ref{sec:background}, we shortly recapitulate the MFT and the calculation of current fluctuations. The AP assumption---as presented in~\cite{Bodineau2004}---is outlined, and a sufficient and necessary condition for its validity is derived for continuous transitions, similarly to~\cite{Shpielberg2016}.  Sec.~\ref{sec:models} reviews the models we probe numerically for DPTs. In Sec.~\ref{sec:results}, we present evidence for dynamical phase transitions  for each of these models. Sec.~\ref{sec:summary} summarizes our findings, and presents further directions of study.

% ======================================================================================== %

\section{Theoretical Background
\label{sec:background}}

We consider diffusive particles in a one-dimensional system of size $L$ as described by the MFT (see~\cite{Derrida2007,Imparato2009} for a microscopic derivation). Particles can be injected to---or extracted from---the system at the boundaries only. This condition is expressed by the continuity equation,
\begin{equation}
\partial_\tau \varrho \left(x,\tau\right)=-\partial_x j\left(x,\tau\right),
\label{eq:continuity equation}
\end{equation}
which relates the coarse grained density $\varrho \left(x,\tau\right)$  and current density $j\left(x,\tau\right)$ by using rescaled coordinates for space $x \in \left[ 0,1 \right]$ and time $\tau \in \left[ 0,t/L^2 \right] $.

The MFT states that the probability to observe a fluctuation $\left(\varrho,j\right)$ can be written using only the macroscopic diffusion coefficient $D$ and conductivity $\sigma$, which generally depend on the local density $\varrho$. 
To define $D\left(\varrho\right)$ and $\sigma\left(\varrho\right)$, we consider a system coupled at each end to a reservoir with fixed densities $\varrho_L$  and $\varrho_R$ at $x=0$ and $x=1$, respectively. We take $\varrho_L=\varrho$  and  $\varrho_R=\varrho+\Delta \varrho  $ with $\Delta\varrho\ll 1$. The diffusion and conductivity are defined using the first two cumulant coefficients of the integrated number of particles $Q$,   
\begin{subequations}
\label{eq:Two cumulants}
\begin{align}
J_s \equiv \frac{\mean{Q}}{t} & =  -\frac{1}{L} \, D\left(\varrho\right) \Delta\varrho, \\
\frac{\mean{Q^2}_C}{t} & = \frac{1}{L} \sigma\left(\varrho\right),
\end{align} 
\end{subequations}
where $\mean{\,\cdot\,}$ stands for an averaging with respect to the steady state probability distribution, and $\mean{Q^2}_C \equiv \mean{Q^2}-\mean{Q}^2$.

The large deviation principle assumes that the probability to observe a net transfer of $Q$ particles is given by   
\begin{equation}
P_t\left(Q\right) \sim \exp\left[ -t \, \Phi\left(J=Q/t\right)\right],
\end{equation}
where $J$ is the mean constrained current and $\Phi\left(J\right)$ is the large deviation function. Using the MFT,  $\Phi\left(J\right)$  is expressed as a minimization problem~\cite{Bertini2005,Bertini2006,Shpielberg2016}. Obtaining  $\Phi\left(J\right)$  explicitly requires finding an optimal fluctuation $\left(\varrho,j\right)$  that satisfies Eq.~\eqref{eq:continuity equation} and the mean constrained current $J=Q/t$. 

Another useful representation of current statistics is obtained from the cumulant generating function $\mu\left(\lambda\right)$. Here, the two previous constraints are relaxed, and the mean constrained current $J$  is replaced by $\lambda$,  and a Lagrange multiplier $p$ is introduced to account for the continuity equation \eqref{eq:continuity equation}.  The cumulant generating function is the Legendre transform of the large deviation function, $\mu\left(\lambda\right) = \sup_{J} \left( \Phi\left(J\right)-\lambda J \right)$.

The cumulant generating function is then explicitly written as a minimization problem~\cite{Imparato2009,Jordan2004,Tailleur07} , %
\begin{equation}
\mu \left(\lambda\right) = \min_{\varrho,p} \int \D x \D\tau \, \left( p\, \partial_\tau \varrho  - \mathcal{H} \right), 
\label{eq: CGF explicit}
\end{equation}
where~\footnote{Here we assume the number of particles in the system is bounded.}
\begin{equation}
\mathcal{H} = - D\left(\varrho\right)\partial_x \varrho \, \partial_x p 
              + \tfrac{1}{2} \sigma\left(\varrho\right) (\partial_x p)^2.
\medskip{}
\end{equation}
The trajectories $\left(\varrho,p\right)$, which solve the minimization problem, satisfy Hamilton equations subjected to the boundary conditions~\cite{Shpielberg2016}
\begin{subequations}
\label{eq:BC-notAP}
\begin{align}
\varrho\left(x=0,\tau\right)  & = \varrho_L &
\varrho\left(x=1,\tau \right) & = \varrho_R \\
 p \left(x=0,\tau\right)      & = 0 &
 p \left(x=1,\tau\right)      & = -\lambda.
\end{align}
\end{subequations}

Solving this problem proves generally difficult. In~\cite{Bodineau2004}, it was conjectured (the AP assumption) that the optimal density profile is time-independent~\footnote{We will also consider $p$ to be time-independent, although it may not be necessarily so.}. In this case, the AP assumption implies that $\left( \varrho_0\left(x\right),p_0\left(x\right)\right)$  are solutions of
\begin{subequations}
\label{eq:AP Hamilton eq}
\begin{align}
 \partial_x \left( D_0 \, \partial_x \varrho_0 - \sigma_0 \, \partial_x p_0 \right) & = 0  \\
 -D_0 \, \partial_{xx} p_0 - \tfrac{1}{2} \sigma'_0 \left( \partial_x p_0 \right)^2 & = 0  ,
\end{align} 
\end{subequations}
where $\sigma'\left(\varrho\right)\equiv{\D\sigma}/{\D\varrho}$ and $D_0,\sigma_0,\sigma'_0$ are evaluated at $\varrho_0$. According to~\eqref{eq:BC-notAP}, the boundary conditions for Eqs.~\eqref{eq:AP Hamilton eq} are
\begin{subequations}
\label{eq:BC}
\begin{align}
 \varrho_0\left(x=0\right) & = \varrho_L &
 \varrho_0\left(x=1\right) & = \varrho_R \\
   p_0 \left(x=0\right)    & = 0 &
   p_0 \left(x=1\right)    & = -\lambda.
\end{align}
\end{subequations}

In~\cite{Shpielberg2016}, a sufficient and necessary condition has been derived, which verifies whether or not the AP solution is a locally minimal solution. It relies on showing that only if there is an allowed fluctuation  $\left( \delta \varrho , \delta p\right)$  about the AP solution $\left(\varrho_0, p_0 \right)$  that gives a lower value to $\mu\left(\lambda\right)$ in Eq.~\eqref{eq: CGF explicit}, then the AP solution is incorrect (this approach disregards first order DPTs)~\footnote{This scheme can only probe small fluctuations about the AP solution. To our knowledge, there are no sufficient and necessary conditions which accommodate large fluctuations.}. 
It was found that the AP assumption is not valid for  $\mu \left( \bar{\lambda}\right)$ if and only if~\cite{Shpielberg2016}
\begin{widetext}
\noindent (a) $\exists \left( \bar{\omega},\bar{\lambda} \right)$ such that there exists a non-trivial solution to the equations 
\begin{subequations}
\label{eq:fluc f,g}
\begin{align}
\I \omega f_\omega  & = \partial_x \Bigl[   D'_0 \left( \partial_x \varrho_0 \right)  f_\omega 
                                          + D_0  \left( \partial_x f_\omega \right)
                                          - \sigma'_0 \left( \partial_x p_0 \right)  f_\omega 
                                          - \sigma_0  \left( \partial_x g_\omega \right)
                                    \Bigr] \\
\I \omega g_\omega  & = \Bigl[ - D'_0 \, \partial_{xx} p_0 
                         - \tfrac{1}{2}\sigma''_0 \left(\partial_x p_0 \right)^2 \Bigr]  f_\omega
                         - D_0 \left(\partial_{xx} g_\omega\right) 
                         -  \sigma'_0 \left( \partial_x p_0 \right) \left( \partial_x g_\omega \right);
\end{align}
\end{subequations}	
(b) for these $ \left( \bar{\omega},\bar{\lambda} \right)$, $\delta s_{\bar{\omega}}^2<0$, where
\begin{equation}
\label{eq:delta s}
\delta s_\omega ^2 = \int \D x \, \biggl[ 
     \frac{D'_0\sigma'_0-D_0\sigma''_0}{4D_0} \left( \partial_x p_0\right)^2 \left|f_\omega\right|^2 
   + \frac{\sigma_0}{2} \left|\partial_x g_\omega\right|^2 \biggr].
\end{equation}
\end{widetext}
Here $\left(  f_\omega\left(x\right) , g_\omega\left(x\right) \right)$  are the Fourier modes of the fluctuations  $\left(\delta \varrho , \delta p\right)$ namely, $\delta\varrho = \sum_\omega f_\omega \E^{\I\omega\tau}$ and $\delta p = \sum_\omega g_\omega \E^{\I\omega\tau}$, 
and $\delta s_{\omega}^2$ is the Fourier transform of the variation of $\mu \left( \lambda \right)$ to second order in $\delta\varrho ,\, \delta p$. 
The boundary conditions for $\left(f_\omega,g_\omega\right)$ are 
\begin{subequations}
\label{eq:fg-BC}
\begin{align}
f_\omega\left(x=0\right)   = 
f_\omega\left(x=1\right) & = 0 \\
g_\omega\left(x=0\right)   = 
g_\omega\left(x=1\right) & = 0,
\end{align}
\end{subequations}
since the solutions  $\left(\varrho_0 + \delta \varrho , p_0 + \delta p\right)$ must satisfy the boundary conditions~\eqref{eq:BC}. 

Note that  $f_\omega=g_\omega = 0$ is a trivial solution to the linear Eqs.~\eqref{eq:fluc f,g}.  Clearly that trivial solution never corresponds to a DPT as it  yields $\delta s_\omega ^2 = 0$. Therefore, the main task of finding a numerical solution of Eqs.~\eqref{eq:fluc f,g} is to force out that trivial solution and to find nontrivial ones.

We note that in~\cite{Baek2016b}, it was necessary to develop a non-linear perturbation theory, namely, to include higher order terms, such as $\delta \varrho^2, \delta p^2, \delta \varrho \, \delta p$. However, it seems that the method presented in~\cite{Baek2016b} cannot be extended to find time-dependent transitions. For the models considered here, linear perturbation theory is sufficient to observe DPTs.

There is no direct method to systematically search for first order DPTs. However, one may encounter them while solving the AP equations~\eqref{eq:AP Hamilton eq} with boundary conditions~\eqref{eq:BC}. It happens when either the large deviation function $\Phi \left( J \right)$ or the cumulant generating function $\mu \left( \lambda \right)$ favor one solution below some critical value and a second one above it (see Appendix~\ref{sec:app:Order-of-transition}). There is no guarantee, though, that there is no third solution, which is always favorable.

% ======================================================================================== %

\section{Models
\label{sec:models}}

As for now, there is no known example of a process leading to a time-dependent second order DPT under boundary drive. Note that from Eq.~\eqref{eq:delta s}, one can infer a sufficient condition for the validity of the AP solution~\cite{Shpielberg2016}, namely, 
\begin{equation}
D'_0\sigma'_0\geq D_0\sigma''_0.
\label{eq:suff}
\end{equation}

Several models, \eg the Symmetric Simple Exclusion Process (SSEP)~\cite{Derrida2004,Bodineau2004} and the Zero Range  Process~\cite{Bodineau2004,Bertini2006},~\footnote{Where it is assumed there is no accumulation of particles in the system. See~\cite{Hirschberg2015} for an example where this assumption does not hold.} satisfy Eq.~\eqref{eq:suff}, so that the AP can be used to evaluate the cumulant generating function.  

Here we present three models and discuss the occurrence of a DPT:  (\hyperref[subsec:Mexican]{A}) a toy model with three extremal points for $\sigma\left(\varrho\right)$, (\hyperref[subsec:KMP]{B}) the KMP model, and (\hyperref[subsec:LongRangeHop]{C}) the Long-Range Hopping with Exclusion model suggested by Bodineau~\cite{BodineauPrivate}. We will analyze them numerically in the next section.

% ===== D and sigma Figure ===== % 
\begin{figure}[ht]
\begin{centering}
\includegraphics[width=1\columnwidth]{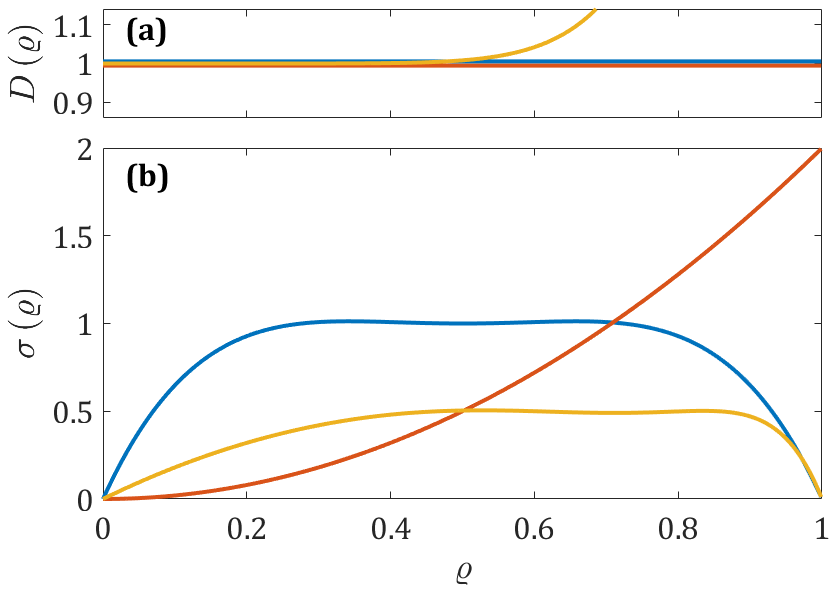}
\par\end{centering}
\caption{\label{fig:D and sigma} 
    (a) The diffusion $D$, and (b) the conductivity $\sigma$. The models under inspection: in blue -- the Mexican Flat Hat model for $A=1$ and $B=-20$; in red -- the KMP model; in yellow -- the Long-Range Hopping with Exclusion model for $\alpha=1/24$ and $\beta=9$. }
\end{figure}

\subsection{A Toy Model -- the Mexican Flat Hat
\label{subsec:Mexican}}
The first model we consider is built in an attempt to find a boundary driven model which presents first and second order DPTs  as well as being relatively simple to analyze numerically. The Mexican Flat Hat is not derived from any microscopic dynamics. We take $D\left(\varrho\right)=1$ with $\sigma\left(\varrho\right) = A\left(\varrho-\frac{1}{2}\right)^2 + B\left(\varrho-\frac{1}{2}\right)^4 - \frac{A}{4} - \frac{B}{16} $, such that we may have a region in $\varrho$, where the sufficient condition in Eq.~\eqref{eq:suff} is not fulfilled. In that region, we look for a violation of the AP.

\subsection{The KMP Model
\label{subsec:KMP}}
The KMP model~\cite{Kipnis1982} was the first model of heat transfer shown to satisfy Fourier's law -- Fick's law counterpart for heat transfer. 

In the KMP model, each site $i\in 1,\ldots,L$  stores an energy $e_i\geq 0$. At each time step $t$, we choose two neighboring sites $i$ and $i'$. They redistribute their respective energies according to a random value $p \in \left[0,1\right]$, namely, $e_i \left(t+\D t\right) = p\,e_i\left(t\right) + \left(1-p\right) e_{i'}\left(t\right)$, and $e_{i'}\left(t+\D t\right) = \left(1-p\right) e_i\left(t\right) + p\,e_i'\left(t\right)$. This implies energy conservation at each time step, with fast equilibration between neighboring sites. The boundaries are considered as fictitious sites with energies drawn from a Boltzmann distribution of respective temperatures $\varrho_l$ and $\varrho_r$. 

Macroscopically, the KMP model is obtained by taking $D\left(\varrho\right)=1$ and $\sigma\left(\varrho\right)=2\varrho^2$~\cite{Kipnis1982}, where $\varrho\left(x,\tau\right)$ denotes the energy density instead of the particle density.

\subsection{Long-Range Hopping with Exclusion Model
\label{subsec:LongRangeHop}}
The long range hopping exclusion model proposed by Bodineau~\cite{BodineauPrivate} is a one dimensional lattice gas model with $L$ sites whose occupancy $n_i=\left\{0,1\right\}$  for $i\in 1,\ldots,L$. A particle can hop from site $i$ to a nearest neighbor site $i \pm 1$ with rate $1$ just like in the SSEP. 
Unlike SSEP, however, a particle  is also allowed to hop from site $i$ to site $i \pm (\beta + 1)$ with a rate $\alpha$ provided that the $\beta$ sites separating them are all occupied, as depicted below
%
% * A depiction of the LHE model. *
%
\begin{equation*}
\xy
\xymatrix "M" @C=0.5em {
\fullsite & \emptysite 
& \fullsite \ar`u[rrrrr]`[rrrrr]^{\displaystyle\alpha}[rrrrr] \ar`u[l]`[l]_{\displaystyle 1}[l]
& \fullsite & \fullsite & \cdots & \fullsite & \emptysite & \emptysite
}%
\POS"M1,4"."M1,7"!C*\frm{_\}},+D*++!U\txt{$\beta$ occupied sites}
\endxy
\end{equation*}

This model is a gradient model, a result which allows to obtain $D$ and $\sigma$ analytically~\cite{Spohn1992Part2Chap2}.  We obtain $D\left(\varrho\right) = 1 + \alpha\left(\beta + 1\right)^2 \varrho^\beta$ and $\sigma\left(\varrho\right) = 2\varrho\left(1-\varrho\right) D\left(\varrho\right)$ with $\varrho \in \left[ 0,1\right]$. This model allows some freedom in the form of  $D$ and $\sigma$  due to the free parameters  $1\geq\alpha\geq 0$  and $\beta \in \mathbb{N}$.

% ======================================================================================== %

\section{Numerical Method -- Results
\label{sec:results}}

To probe for a  second order DPT, we solve Eqs.~\eqref{eq:AP Hamilton eq} and~\eqref{eq:fluc f,g} with the boundary conditions~\eqref{eq:BC} and~\eqref{eq:fg-BC}. 

A na\"{\i}ve attempt to find a solution of Eqs.~\eqref{eq:fluc f,g} using the numerical solution of Eqs.~\eqref{eq:AP Hamilton eq} will \emph{almost always} yield the trivial solution $f_\omega=g_\omega=0$ due to the linearity of the equations and their corresponding boundary conditions~\eqref{eq:fg-BC}.  We therefore employ a ``sniping method'': we condition the numerical solver to the boundary conditions
\begin{subequations}
\label{eq:fg-BC-new}
\begin{align}
f_\omega\left(x=0\right) & = 0 &
f_\omega\left(x=1\right) & = 0 \\
g_\omega\left(x=0\right) & = 0 &
\partial_x g_\omega\left(x=0\right) & = 1,
\end{align}
\end{subequations}
thus forcing the system away from the trivial solution. Note that due to the linearity of Eqs.~\eqref{eq:fluc f,g}, the particular choice of the value for the derivative condition, $\partial_x g_\omega\left(0\right)=1$,  changes the solution $\left(f_\omega,g_\omega\right)$ only by a multiplicative factor, which is innocuous to the problem at hand. Generally, for a given $\left(\lambda,\omega\right)$  the solution of Eqs.~\eqref{eq:fluc f,g} with the boundary conditions~\eqref{eq:fg-BC-new} does not satisfy the original boundary conditions~\eqref{eq:fg-BC}. We  identify a \emph{proper} solution  $\left(f_\omega,g_\omega\right)$ only if  $\left|g_\omega\left(1\right)\right|=0$.

By changing the boundary conditions from~\eqref{eq:fg-BC} to~\eqref{eq:fg-BC-new}, we find proper solutions by systematically scanning the $\left(\lambda,\omega\right)$ space. We then check the value of    $\delta s_\omega^2 $. Only  proper solutions with  $\delta s_\omega^2<0$  indicate a second order DPT. 

We note that, given a solution at some $\lambda_s$, one must check that solutions also exist in a finite range around it. Otherwise, it is required to go beyond the linear perturbation theory considered here~\cite{Baek2016b}.

% ====== Between Extrema Depiction ===== %
\begin{figure}[ht]
\begin{centering}
\includegraphics[width=0.75\columnwidth]{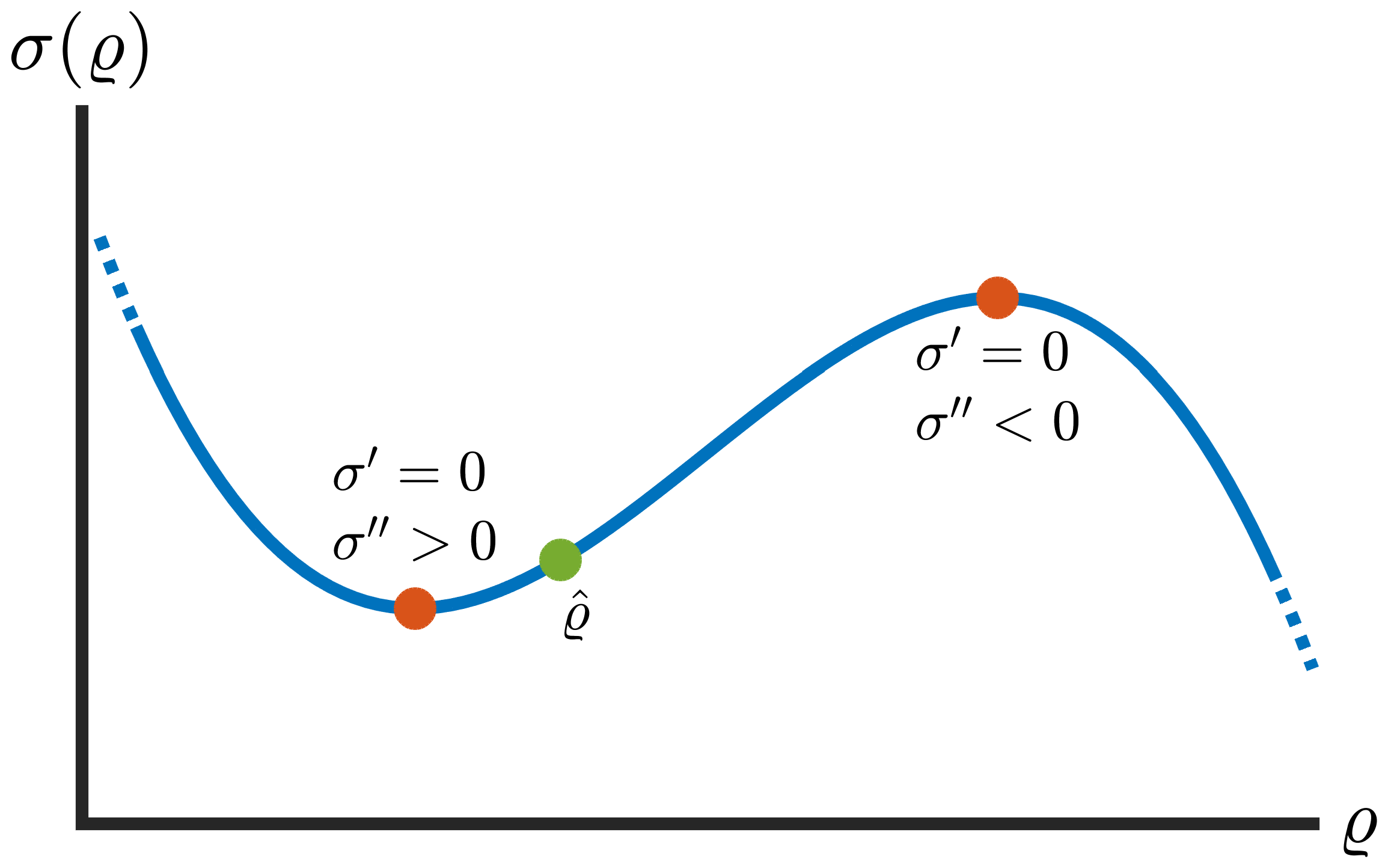}
\par\end{centering}
\caption{\label{fig:between_extrema}
Fixing the boundary conditions $\varrho_L=\varrho_R=\hat{\varrho}$ between a minimal and a maximal point in $\sigma$. For $D=1$, we expect the AP density profile to increase as the current is increasing (see Appendix~\ref{sec:app:LDF intuition} for some intuition). However, at some point the density profile  reaches a maximal value of $\sigma$. Further heightening of the density profile is no longer favorable.  This calls for a different characteristic  solution of the density profile.  It could conceivably be a different AP solution or a time-dependent solution. 
}
\end{figure}

\subsection{A Toy Model -- the Mexican Flat Hat}
The motivation  behind  this  model is to lure the AP solution to follow a favorable path of the density profile due to a rapid increase of $\sigma$. This trend becomes  unfavorable, if the density profile hits a maximal point of  $\sigma$ (see Fig.~\ref{fig:between_extrema} and Appendix~\ref{sec:app:LDF intuition}).  In order to achieve that, we consider a model where $\sigma$ has at least one minimum and one maximum, and set $\varrho_L=\varrho_R = \hat{\varrho}$ such that  $\hat{\varrho}$ lies between the two extreme points (see Fig.~\ref{fig:between_extrema}). 

From the large deviation function picture (see Appendix~\ref{sec:app:LDF intuition}), one expects that increasing $J$ (correspondingly, $\lambda$),  manifests in an increase of the density profile, since  $\sigma$  is monotonously increasing initially for $\varrho>\hat{\varrho}$. However, for a sufficiently large current, the density profile reaches the maximal value in $\sigma$. Then, it is no longer beneficial to continue and increase the density profile. At this point a different solution, although not necessarily time-dependent, should take over. 

\begin{figure}[t]
\begin{centering}
\includegraphics[width=1\columnwidth]{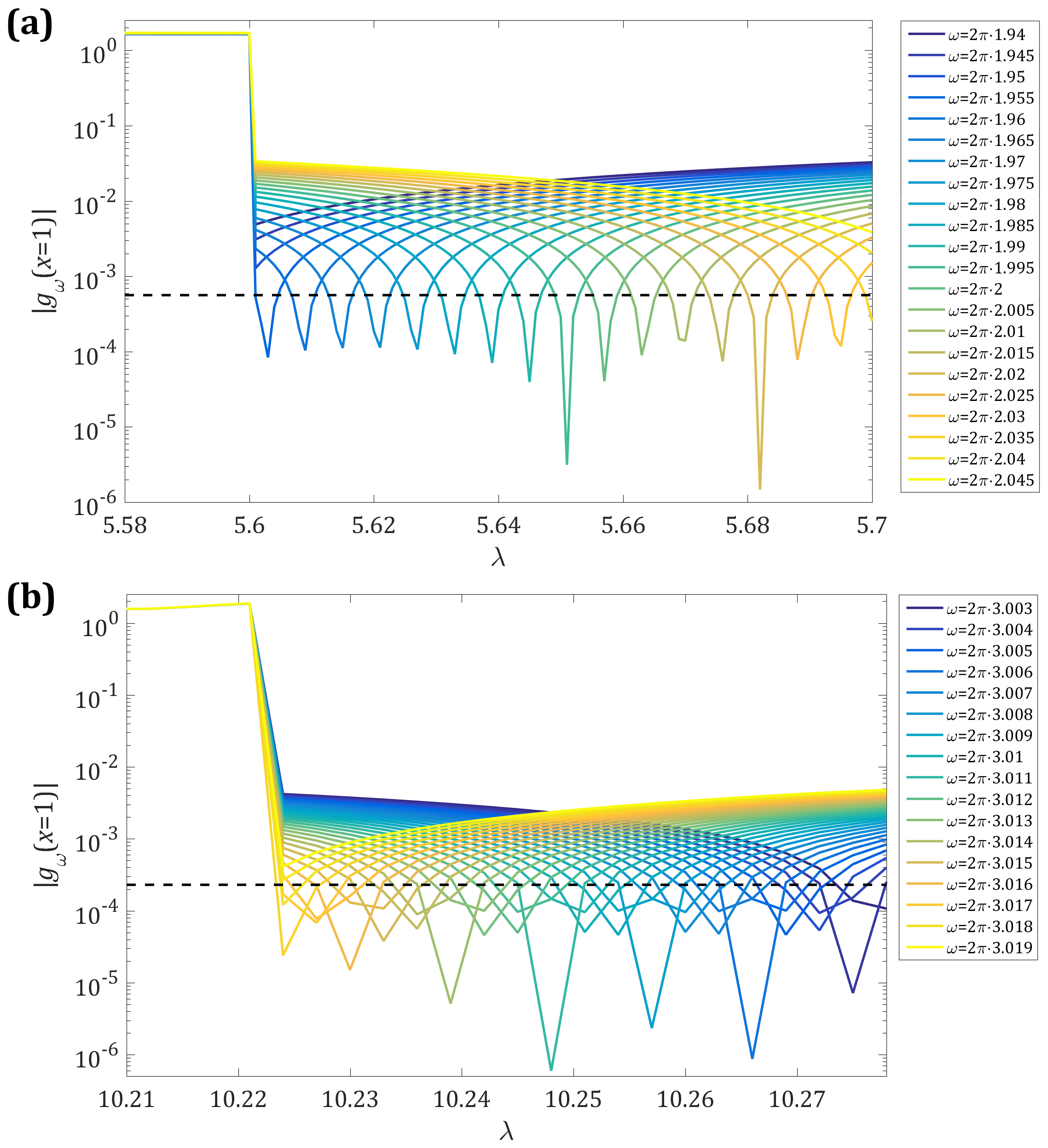}
\par\end{centering}
\caption{\label{fig:fluc_Both}
The sniping method results for the Mexican Flat Hat and the Long-Range Hopping with Exclusion models.
We are interested in finding a solution for which $\left|g_\omega\left(x=1\right)\right|=0 $ for some $\lambda$ and $\omega$ using the sniping method. Identification of numerical zeroes is made by going below the numerical error bars implying the existence of a  solution to \eqref{eq:fluc f,g} with the boundary conditions \eqref{eq:fg-BC}. 
Full lines: the absolute values of $g_\omega\left(x=1\right) $ using the sniping method. Dashed lines: the numerically estimated errors on  $\left|g_\omega\left(x=1\right)\right|$. 
(a) The Mexican Flat Hat model for $A=1$ and $B=-20$. 
(b) The Long-Range Hopping with Exclusion model for $\alpha=1/24$ and $\beta=9$.
No solutions to \eqref{eq:fluc f,g} were found below $\lambda=5.6$ in (a) and below $\lambda=10.22$ in (b).
}
\end{figure}

\begin{figure}[t]
\begin{centering}
\includegraphics[width=1\columnwidth]{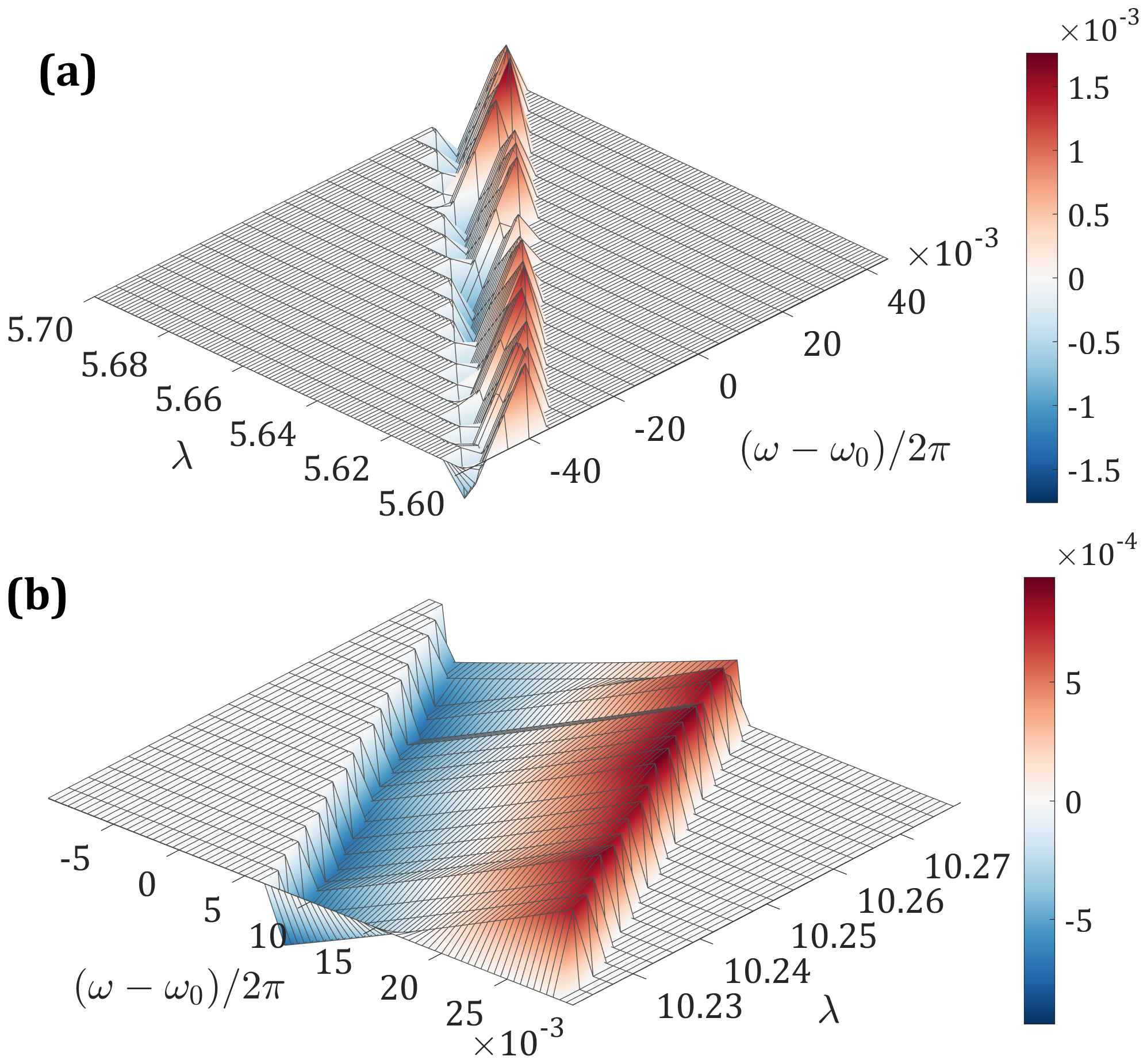}
\par\end{centering}
\caption{\label{fig:ds2_Both} 
The value of $\delta s^2_\omega$  for a range of  $\omega$ and $\lambda$  above the critical region for the Mexican Flat Hat and the Long-Range with Exclusion models. Here, the red hues represent the positive values of $\delta s^2_\omega$, and blue hues the negative values. 
The locations for which $\delta s^2_\omega=0$ imply that only the trivial solution $f_\omega=g_\omega=0$ exists.
The negative values of $\delta s^2_\omega$ imply a second order DPT. 
(a) The Mexican Flat Hat model for $\omega_0 = 4\pi$.
(b) The Long-Range Hopping with Exclusion model for $\omega_0 = 6\pi$.
The reference points $\omega_0$ in (a) and in (b) are chosen for \aesthetic reasons. 
}
\end{figure}

We find this behavior to be conceptually correct (see Appendix~\ref{sec:app:LDF intuition} for details). First, we notice a discontinuous transition between two AP solutions, a concave density profile, and a  convex density profile at $\lambda_{c_1} $. Namely, for  $0<\lambda<\lambda_{c_1}$ the concave density profile has a lower value for $\mu \left( \lambda\right)$ while for  $\lambda_{c_1}<\lambda<\lambda_{c_2}$ the convex density profile is favorable). After which, at a higher $\lambda_{c_2}$, the convex density profile becomes unstable to small time-dependent fluctuations.  We find this behavior for $\hat{\varrho}=0.55$, with $\lambda_{c_1}\approx3.70$ and $\lambda_{c_2}\approx5.60$, which correspond to $J_{c_1}\approx4.91$ and $J_{c_2}\approx8.41$. For the second order DPT at $\lambda_{c_2}$,  there is a non-zero mode near $\omega_0 = 4\pi $ that allows non-trivial fluctuations  $\left(  f_\omega , g_\omega \right)$ (see Fig.~\hyperref[fig:fluc_Both]{\ref*{fig:fluc_Both}a}). This fluctuation was found to give a lower value than the AP solution for the cumulant generating function $\mu\left(\lambda\right)$ as seen in Fig.~\hyperref[fig:ds2_Both]{\ref*{fig:ds2_Both}a}. We note that this behavior can also be found for non-equal boundary conditions.

% ====== Sniping Method ===== %

\subsection{The KMP Model}
The KMP model was extensively studied numerically in the context of current fluctuations. For periodic boundary conditions, it was found that there exist such DPTs, and that for a certain $\lambda$ the optimal solution becomes a traveling wave rather than a fixed density profile. However, for the boundary driven case, there is no theoretical or numerical indication for such DPTs. In~\cite{Hurtado2009} the cumulant generating function was probed for a range of values of $\lambda$ using an exact simulation of the dynamics. For reservoir values of $\varrho_L=1$ and $\varrho_R=2$, they were able to verify the AP solution up to $\lambda \approx [-0.7,0.35]$, which corresponds to the range $J \approx [-3.5,6.64]$. Using the sniping method, we are able to verify that the AP solution is valid in the range $ \lambda \approx [-0.9977,0.49]$, which corresponds to $J \approx [-78.8,30.66]$. Note that for the above boundary conditions, $\lambda \in \left(-1,0.5\right)$ corresponds to the whole range of current fluctuations. We find no non-trivial solutions to Eqs.~\eqref{eq:fluc f,g} for these boundary conditions as well as for various others. 

In summary, we were able to show that the AP solution is a minimum solution for very large currents---an order of magnitude improvement as compared to previous numerical results~\cite{Hurtado2009}.

% ===== Fluctuation Value ===== %

\subsection{Long-Range Hopping with Exclusion Model}
Contrary to the Mexican Flat Hat model, the Long-Range Hopping with Exclusion model is derived from a microscopic model. 
We produce an almost constant range of $D$ by applying $\alpha=1/24$ and $\beta=9$, as shown in Fig.~\ref{fig:D and sigma}, and get $\sigma$ having three extremal points much in the spirit of the Mexican Flat Hat model. Probing for boundary conditions of $\varrho_L=\varrho_R=0.75$,  we find the same behavior as in the Mexican Flat Hat model, and with $\lambda_{c_1}\approx5.01$ and $\lambda_{c_2}\approx10.22$, which correspond to  $J_{c_1}\approx3.60$ and $J_{c_2}\approx7.54$. At the continuous transition at $\lambda_{c_2}$,  there is a non-zero mode near $\omega_0 = 6\pi $ that allows non-trivial fluctuations  $\left(  f_\omega , g_\omega \right)$ (see Figs.~\hyperref[fig:fluc_Both]{\ref*{fig:fluc_Both}b} and~\hyperref[fig:ds2_Both]{\ref*{fig:ds2_Both}b}). We note that the description of convex and concave solutions (see Appendix~\ref{sec:app:Order-of-transition}) for this model is oversimplified due to the nontrivial $D\left(\varrho\right)$. Beyond $\lambda_{c_1}$ the AP density profile can support more than one  point at which $\frac{\D\varrho}{\D x}=0$.

% ======================================================================================== %

\section{Summary
\label{sec:summary}}

We have presented a first evidence of two types of DPTs in the context of current fluctuations in boundary driven systems. The first being a discontinuous transition between two different AP solutions, and the second being a continuous transition from an AP solution to a time-dependent solution. We have also numerically verified that the KMP model does not break the AP assumption under small perturbations up to  high currents. It is to be understood that a key ingredient in observing these DPTs is to  set the boundary conditions such that the steady state density profile is in the regime between two extreme values of the conductivity $\sigma$.  We note that this scheme needs not to be unique, and does not guarantee DPTs for general models. Moreover, continuous and discontinuous DPTs do not nessesarily come in pairs.

One open question is to characterize the role the diffusion coefficient $D$ plays in such transition.  We have also been unable to find a simple model for which the two types of DPTs can be analytically shown to occur.  It is evident that there is a significant lack of understanding of the typical time-dependent density profile for boundary driven processes, as opposed to periodic boundary conditions, where after the transition the density profile behaves like a traveling wave~\cite{Zarfaty2015a}. Moreover, it is of interest, although inherently difficult~\cite{Garnier2005},  to probe this transition in some experimental realization.

% ======================================================================================== %

\begin{acknowledgments}
This work has been supported by ANR-14-CE25-0003 and by Israel Science Foundation Grant No. 924/09.
We would like to thank Yongjoo Baek for bringing to our attention his work, and to Thierry Bodineau for suggesting his model. We acknowledge Kirone Mallick, Avi Aminov, and Vivien Lecomte for useful discussions. 
\end{acknowledgments}

% ======================================================================================== %

% a small hack to start the appendix leveled
\newpage
\begin{widetext}
\end{widetext}

% === Appendix === %

\appendix

% ======================================================================================== %

\section{The Large Deviation Function Formalism
\label{sec:app:LDF intuition}}

The purpose of this section is to present the large deviation function approach to current statistics, and to provide an intuitive approach to search for  multiple AP solutions. 
The MFT provides a formal expression to the large deviation function (alternative to the cumulant genrating function $\mu \left( \lambda \right)$)
\begin{equation}
\Phi \left(   J \right) =    \min_{\varrho,j\in \mathcal{A}_J}   \frac{L}{t}  \int \D  x \D \tau \, \mathcal{L} \left(  \varrho, j  \right), 
\end{equation}
where 
\begin{equation}
\mathcal{L}\left(  \varrho, j  \right) = \frac{\left(  j+D\partial_x \varrho  \right)^2}{2\sigma},
\end{equation}
such that $\varrho	\left(x=0,\tau \right) = \varrho_L$,  $\varrho \left(x=1,\tau \right) = \varrho_R$, and $\mathcal{A}_J$ is the set of all currents $j \left( x,\tau \right)$ such that $\int \D x \D \tau \, j \left( x,\tau \right) = Jt/L^2 $.  The AP assumption gives an upper bound $U \left( J \right)$ to $\Phi \left( J \right ) $, where 
\begin{equation}
U \left( J \right) = \frac{1}{L}  \min_{\varrho \left( x  \right)}  \int \D x \: \mathcal{L}_J  \left( \varrho \left(x \right) \right)    
\end{equation}
with $\mathcal{L}_J =\frac{\left(  J+D\partial_x \varrho  \right)^2}{2\sigma} $,  $\varrho	\left(x=0\right) = \varrho_L$, and $\varrho	\left(x=1\right) = \varrho_R$. Finding the optimal solution $\varrho_0 \left( x \right)$ boils down to solving an Euler-Lagrange equation $\frac{\delta \mathcal{L}_J  }{\delta \varrho}   = \frac{\D}{\D x} \frac{\delta \mathcal{L}_J }{\delta \partial_x\varrho}$, which yields 
\begin{equation}
\label{eq:Euler-Lagrange}
\partial_{xx} \varrho_0 +  \left( \partial_x \varrho_0 \right)^2  \left(  \frac{D'_0}{D_0} - \frac{\sigma'_0}{2\sigma_0}   \right) + J^2  \frac{\sigma'_0}{2D_0 ^2 \sigma_0} = 0 .
\end{equation}
It is clear that the AP density profile solution is indifferent to the sign of $J$, although $U\left(J\right)\neq U\left(-J\right)$ for $\varrho_L \neq \varrho_R$. 

Now, after presenting the large deviation function formalism, it is possible to understand the logic behind searching for DPTs in the scenarios depicted in Fig.~\ref{fig:between_extrema}. For $\varrho_L = \varrho_R = \hat{\varrho}$ at  $J=0$, the AP solution yields $\varrho_0 (x) = \hat{\varrho}$. However, as we increase $J$, it is favorable to increase $\varrho_0 \left( x \right)$  above $\hat{\varrho}$  as $\sigma_0$, the denominator of $U$,  increases as well. However, for a large enough $J$, the density profile reaches the maximal point of $\sigma$, as depicted in Fig.~\ref{fig:between_extrema}. Therefore, the density going above this point decreases $\sigma$, and will not be an optimal solution. We thus expect a change of trend in the optimal solution, where it can be a different AP solution  or result in  a transition to a time-dependent solution. 
We are unable to provide a rigorous proof for the transition or even an estimation of the critical $J$  where we expect it  to occur.  However, the numerical solution for the AP seems to correspond to this prediction of a transition.

% ====== Mexican Density Trajectory ===== %
\begin{figure}[t]
\begin{centering}
\includegraphics[width=1\columnwidth]{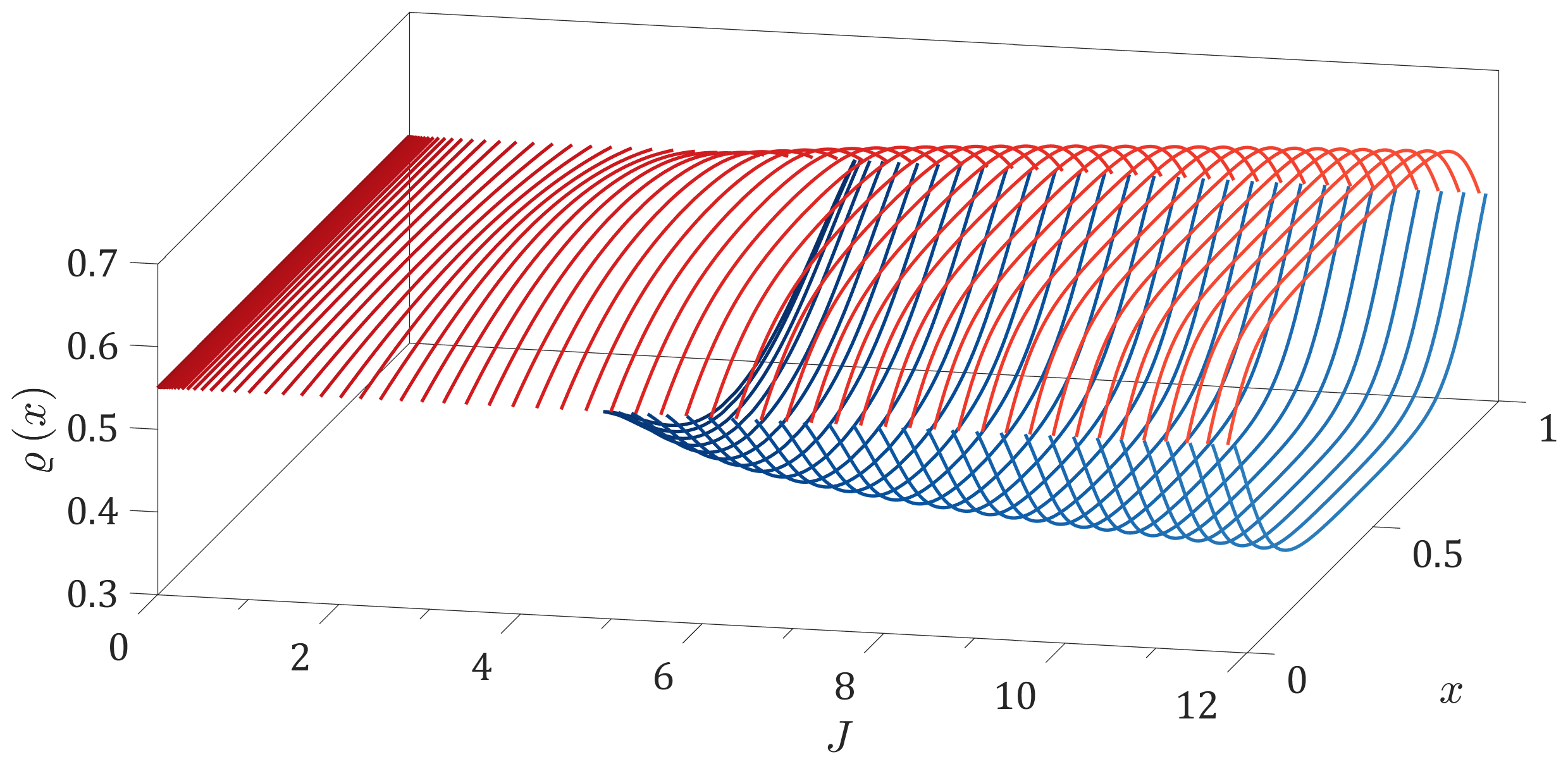}
\par\end{centering}
\caption{\label{fig:density_trajectory}
The trajectories of the density profile $\varrho\left(x\right)$ for different values of $J$. In red the ``frowning'' solutions, and in blue the ``smiling'' solutions.
}
\end{figure}

% ======================================================================================== %

\section{Detecting First Order Transitions
\label{sec:app:Order-of-transition}}

In this Appendix, we  propose  a method to probe for a discontinuous transition between two AP solutions for a specific model with set boundary conditions. This approach heavily relies on insights from~\cite{Bodineau2004}.  We will focus on the Mexican Flat Hat model, but the approach is quite general, albeit harder technically. 

It is possible to show that Eq.~\eqref{eq:Euler-Lagrange} can be reduced to a non-linear first order equation
\begin{equation}
D^2 \left( \varrho \right)  \left( \frac{\D \varrho}{\D x} \right) ^2= J^2 \left( 1+2K\sigma \left( \varrho \right)\right),
\label{eq:quadratic_sol}
\end{equation}
where $K$  is a constant determined by the boundary conditions. For equal boundary conditions $\varrho_L=\varrho_R=\hat{\varrho}$ such that $\sigma' \left( \hat{\varrho} \right) \neq 0$, the density profile is never monotonous, except for $J=0$ where the solution is flat, namely, $\varrho_0 \left( x \right)= \hat{\varrho}$. Since it is not monotonous, there is at least one point for which $\frac{\D \varrho}{\D x}=0$ for differentiable density profiles. It makes sense to consider only a symmetric solution about  $x\mapsto1-x$.   We assume that there is exactly one extreme point in the density profile for any $J\neq 0$, relying on the numerical results obtained by solving Eqs.~\eqref{eq:AP Hamilton eq} and~\eqref{eq:BC} (see Fig.~\ref{fig:density_trajectory}).

% ====== Mexican Conductance Trajectory ===== %
\begin{figure}[t]
\begin{centering}
\includegraphics[width=0.95\columnwidth]{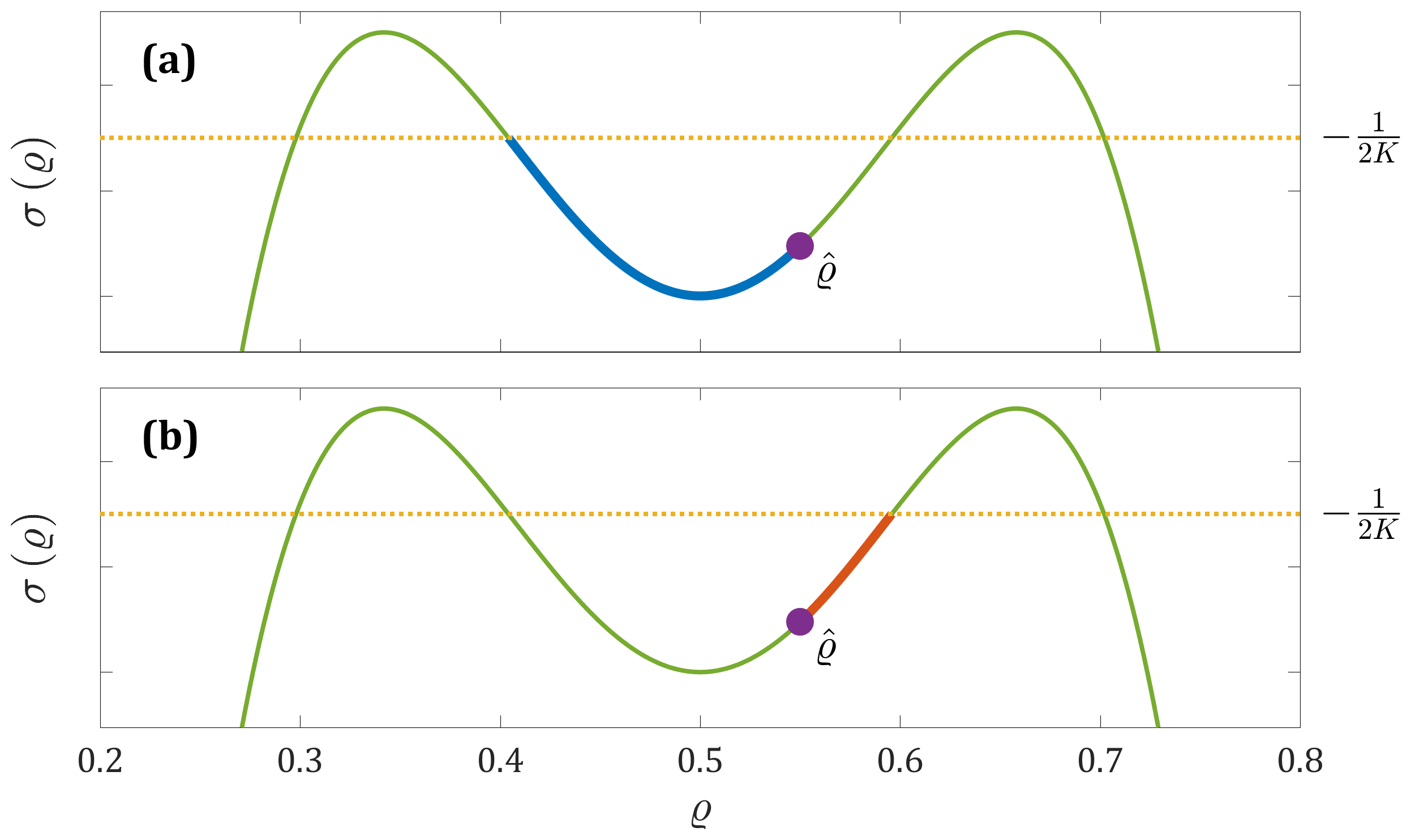}
\par\end{centering}
\caption{\label{fig:conductance_trajectory}
The trajectory of the calculation of $J$. Green line is the conductivity  profile $\sigma$ which plays the part of a potential. Yellow dotted line is the ``total energy''  $-\frac{1}{2K}$. The purple dot represents the boundary value $\varrho_L=\varrho_R=\hat\varrho$. 
(a) The trajectory for the smiling solution (in blue). 
(b) The trajectory for the frowning solution (in red).
}
\end{figure}

The density profile trajectory is analyzed as follows. Since the density profile is non-monotonous, and from~\eqref{eq:quadratic_sol}, we find that    $K \in \left[  -{1}/{2\sigma_{\max}}, -{1}/{2\sigma\left(\hat{\varrho}\right)} \right]$, where $\sigma_{\max}$  is the maximal value $\sigma$ can reach. For a given $K$ in this range, one can find  $\varrho^\star$,  the value  at which the density profile gradient vanishes at $x=1/2$, by using~\eqref{eq:quadratic_sol} to obtain  $\sigma \left( \varrho^\star \right) \leq -\frac{1}{2K}$. Here one can consider  $-\frac{1}{2K}$ playing the role of a total energy, $\sigma \left( \varrho \right)$ being the potential, and $\varrho$ the position. Therefore, one can deduce from this picture the possible density profiles for each $K$ (see Fig.~\ref{fig:conductance_trajectory}).  Since $\sigma$  is a non-monotonous function, much like for a potential picture, there is some degeneracy in the value of $\varrho^\star$.  For each $K$, there are two values for $\varrho^\star$, each corresponding to a different density profile. For  $\varrho^\star > \hat{\varrho}$, the density profile is monotonously increasing in $x\in \left[0,1/2\right)$ and decreasing in $x\in \left(1/2,1\right]$, thus being designated the ``frowning'' solution. For $\varrho^\star < \hat{\varrho}$ instead, the density profile is monotonously decreasing within $x\in \left[0,1/2\right)$ and increasing within $x\in \left(1/2,1\right]$, thus being designated the ``smiling'' solution (see Fig.~\ref{fig:density_trajectory}).

% ====== Mexican LDF Transition ===== %
\begin{figure}[t]
\begin{centering}
\includegraphics[width=0.9\columnwidth]{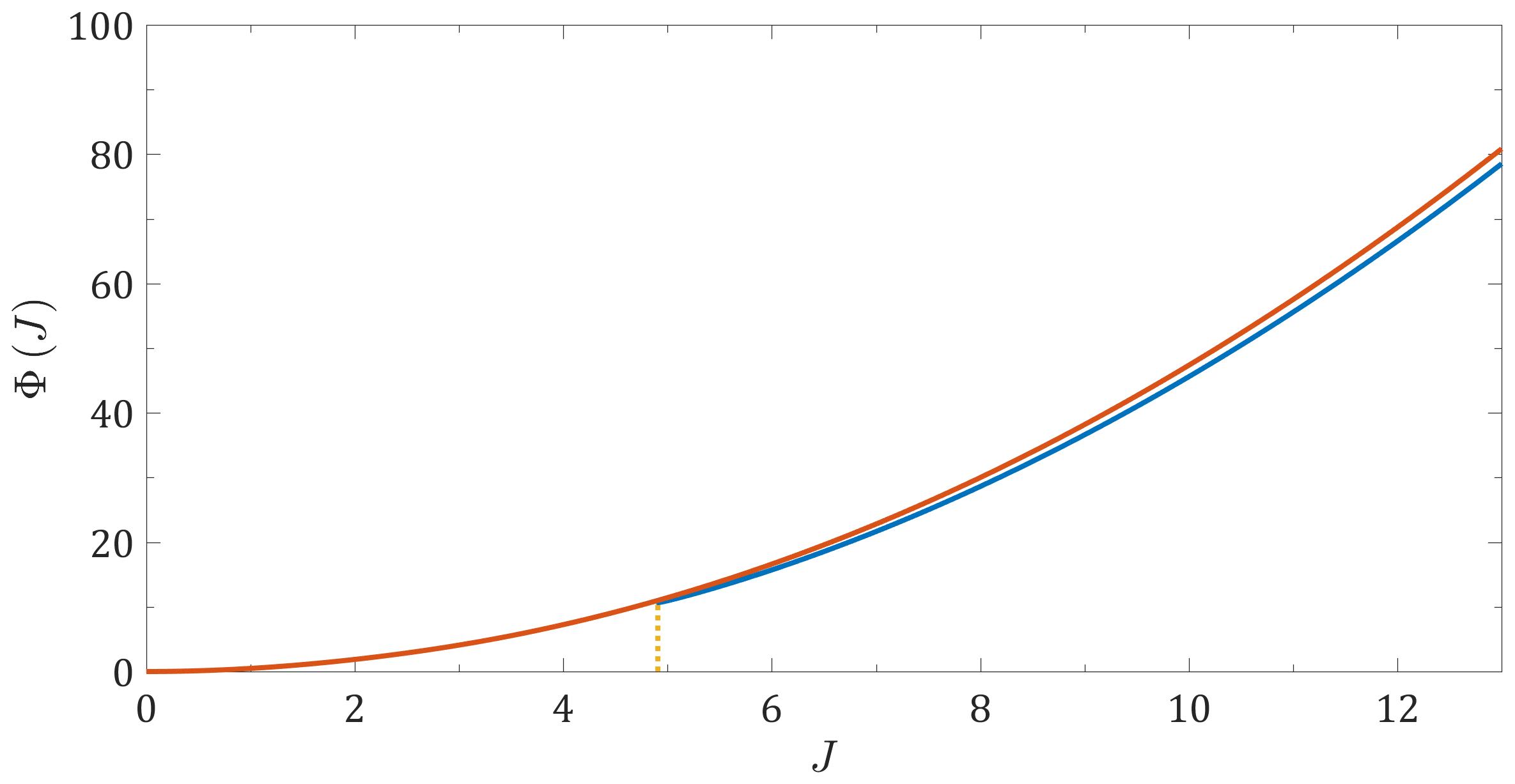}
\par\end{centering}
\caption{\label{fig:LDF_transition}
The large deviation function $\Phi \left( J \right)$ corresponding to the two solutions of $\varrho\left(x\right)$ of Eq.~\eqref{eq:quadratic_sol}. The red line depicts solutions with $\frac{\D\varrho}{\D x}>0$ at $x=0$ (the frowning solution). The blue line depicts solutions with $\frac{\D\varrho}{\D x}<0$ at $x=0$ (the smiling solution). Note that the smiling solution is attainable only from  $J_\mathrm{transition}\simeq4.906$, where it is the favorable solution. The transition is indicated by the dotted yellow line.
}
\end{figure}

Since the sign of the density profile gradient is fixed for $x\in \left[0,1/2\right)$, we can take the square root of~\eqref{eq:quadratic_sol} and obtain 
\begin{equation}
 \frac{\D \varrho}{\D x} = \pm \frac{|J|}{D \left( \varrho \right)} \sqrt{ \left( 1+2K\sigma \left( \varrho \right)\right)},
\label{eq:linear density gradient}
\end{equation}
with $\pm$ for the frowning/smiling solutions, correspondingly (note that the solution is indifferent to the sign of $J$).  From~\eqref{eq:linear density gradient}, one can obtain an integral expression to the current %
\begin{equation}
|J| =  \pm 2 \int^{\hat{\varrho}} _{\varrho^\star} \D \varrho \, \frac{D\left( \varrho\right)}{\sqrt{1+2K\sigma\left( \varrho\right)}} ,
\label{eq:J recover}
\end{equation}
with again $\pm$ for the frowning/smiling solutions, correspondingly. The choice of the $\pm$ sign depends on the $\varrho^\star \gtrless \hat{\varrho}$ case.
 
Using the density profile gradient of Eq.~\eqref{eq:linear density gradient} and the value of the current in Eq.~\eqref{eq:J recover}, we can find the expression for the large deviation function
\begin{equation}
\Phi_\pm \left(J\right) = \pm  2 \int^{\hat{\varrho}} _{\varrho^\star} \D \varrho \,  \frac{D\left( \varrho\right)} {\sigma\left(\varrho\right)} \left[  1 -    \frac{1+K\sigma\left( \varrho\right)}    {\sqrt{1+2K\sigma\left( \varrho\right)}} 
\right],
\end{equation}
where, again, $\pm$ corresponds for the frowning/smiling solutions. Solving numerically and comparing the two AP solutions  for the Mexican Flat Hat  (see Fig.~\ref{fig:LDF_transition}), we observe that for $\hat{\varrho}=0.55$ and for $A=1,B=-20$, there is a single solution (frowning) for low currents as expected. For higher currents where the two solutions exist, where the smiling solution is favorable and marks the onset of a first order DPT as the two solutions are not equal at the transition.

% ======================================================================================== %

% === Bibliography === %

% start the bibliography leveled 
% without \newpage the figures may float incorrectly
\newpage

           % the bibliography file

% \bibliographystyle{apsrev4-1}   % the bibliography style; no article names
%
% \bibliography{AP_breaking}      % the bibliography file

\begin{thebibliography}{55}%
\makeatletter
\providecommand \@ifxundefined [1]{%
 \@ifx{#1\undefined}
}%
\providecommand \@ifnum [1]{%
 \ifnum #1\expandafter \@firstoftwo
 \else \expandafter \@secondoftwo
 \fi
}%
\providecommand \@ifx [1]{%
 \ifx #1\expandafter \@firstoftwo
 \else \expandafter \@secondoftwo
 \fi
}%
\providecommand \natexlab [1]{#1}%
\providecommand \enquote  [1]{``#1''}%
\providecommand \bibnamefont  [1]{#1}%
\providecommand \bibfnamefont [1]{#1}%
\providecommand \citenamefont [1]{#1}%
\providecommand \href@noop [0]{\@secondoftwo}%
\providecommand \href [0]{\begingroup \@sanitize@url \@href}%
\providecommand \@href[1]{\@@startlink{#1}\@@href}%
\providecommand \@@href[1]{\endgroup#1\@@endlink}%
\providecommand \@sanitize@url [0]{\catcode `\\12\catcode `\$12\catcode
  `\&12\catcode `\#12\catcode `\^12\catcode `\_12\catcode `\%12\relax}%
\providecommand \@@startlink[1]{}%
\providecommand \@@endlink[0]{}%
\providecommand \url  [0]{\begingroup\@sanitize@url \@url }%
\providecommand \@url [1]{\endgroup\@href {#1}{\urlprefix }}%
\providecommand \urlprefix  [0]{URL }%
\providecommand \Eprint [0]{\href }%
\providecommand \doibase [0]{http://dx.doi.org/}%
\providecommand \selectlanguage [0]{\@gobble}%
\providecommand \bibinfo  [0]{\@secondoftwo}%
\providecommand \bibfield  [0]{\@secondoftwo}%
\providecommand \translation [1]{[#1]}%
\providecommand \BibitemOpen [0]{}%
\providecommand \bibitemStop [0]{}%
\providecommand \bibitemNoStop [0]{.\EOS\space}%
\providecommand \EOS [0]{\spacefactor3000\relax}%
\providecommand \BibitemShut  [1]{\csname bibitem#1\endcsname}%
\let\auto@bib@innerbib\@empty
%</preamble>
\bibitem [{\citenamefont {Mallick}(2015)}]{Mallick2015}%
  \BibitemOpen
  \bibfield  {author} {\bibinfo {author} {\bibfnamefont {K.}~\bibnamefont
  {Mallick}},\ }\href {\doibase 10.1016/j.physa.2014.07.046} {\bibfield
  {journal} {\bibinfo  {journal} {Physica A}\ }\textbf {\bibinfo {volume}
  {418}},\ \bibinfo {pages} {17} (\bibinfo {year} {2015})}\BibitemShut
  {NoStop}%
\bibitem [{\citenamefont {Bernard}\ and\ \citenamefont
  {Doyon}(2016)}]{Bernard2016}%
  \BibitemOpen
  \bibfield  {author} {\bibinfo {author} {\bibfnamefont {D.}~\bibnamefont
  {Bernard}}\ and\ \bibinfo {author} {\bibfnamefont {B.}~\bibnamefont
  {Doyon}},\ }\href {http://stacks.iop.org/1742-5468/2016/i=6/a=064005}
  {\bibfield  {journal} {\bibinfo  {journal} {J. Stat. Mech. Theor. Exp.}\
  }\textbf {\bibinfo {volume} {2016}},\ \bibinfo {pages} {064005} (\bibinfo
  {year} {2016})}\BibitemShut {NoStop}%
\bibitem [{\citenamefont {Pilgram}\ \emph {et~al.}(2003)\citenamefont
  {Pilgram}, \citenamefont {Jordan}, \citenamefont {Sukhorukov},\ and\
  \citenamefont {B{\"{u}}ttiker}}]{Pilgram2003}%
  \BibitemOpen
  \bibfield  {author} {\bibinfo {author} {\bibfnamefont {S.}~\bibnamefont
  {Pilgram}}, \bibinfo {author} {\bibfnamefont {A.~N.}\ \bibnamefont {Jordan}},
  \bibinfo {author} {\bibfnamefont {E.~V.}\ \bibnamefont {Sukhorukov}}, \ and\
  \bibinfo {author} {\bibfnamefont {M.}~\bibnamefont {B{\"{u}}ttiker}},\ }\href
  {\doibase 10.1103/PhysRevLett.90.206801} {\bibfield  {journal} {\bibinfo
  {journal} {Phys. Rev. Lett.}\ }\textbf {\bibinfo {volume} {90}},\ \bibinfo
  {pages} {206801} (\bibinfo {year} {2003})}\BibitemShut {NoStop}%
\bibitem [{\citenamefont {Karrasch}\ \emph {et~al.}(2013)\citenamefont
  {Karrasch}, \citenamefont {Ilan},\ and\ \citenamefont
  {Moore}}]{Karrasch2013}%
  \BibitemOpen
  \bibfield  {author} {\bibinfo {author} {\bibfnamefont {C.}~\bibnamefont
  {Karrasch}}, \bibinfo {author} {\bibfnamefont {R.}~\bibnamefont {Ilan}}, \
  and\ \bibinfo {author} {\bibfnamefont {J.~E.}\ \bibnamefont {Moore}},\ }\href
  {\doibase 10.1103/PhysRevB.88.195129} {\bibfield  {journal} {\bibinfo
  {journal} {Phys. Rev. B}\ }\textbf {\bibinfo {volume} {88}},\ \bibinfo
  {pages} {1} (\bibinfo {year} {2013})}\BibitemShut {NoStop}%
\bibitem [{\citenamefont {Mendoza-Arenas}\ \emph {et~al.}(2013)\citenamefont
  {Mendoza-Arenas}, \citenamefont {Al-Assam}, \citenamefont {Clark},\ and\
  \citenamefont {Jaksch}}]{Mendoza2013}%
  \BibitemOpen
  \bibfield  {author} {\bibinfo {author} {\bibfnamefont {J.~J.}\ \bibnamefont
  {Mendoza-Arenas}}, \bibinfo {author} {\bibfnamefont {S.}~\bibnamefont
  {Al-Assam}}, \bibinfo {author} {\bibfnamefont {S.~R.}\ \bibnamefont {Clark}},
  \ and\ \bibinfo {author} {\bibfnamefont {D.}~\bibnamefont {Jaksch}},\ }\href
  {http://stacks.iop.org/1742-5468/2013/i=07/a=P07007} {\bibfield  {journal}
  {\bibinfo  {journal} {J. Stat. Mech. Theor. Exp.}\ }\textbf {\bibinfo
  {volume} {2013}},\ \bibinfo {pages} {P07007} (\bibinfo {year}
  {2013})}\BibitemShut {NoStop}%
\bibitem [{\citenamefont {Prosen}\ and\ \citenamefont
  {\v{Z}nidari\v{c}}(2009)}]{Prosen2009}%
  \BibitemOpen
  \bibfield  {author} {\bibinfo {author} {\bibfnamefont {T.}~\bibnamefont
  {Prosen}}\ and\ \bibinfo {author} {\bibfnamefont {M.}~\bibnamefont
  {\v{Z}nidari\v{c}}},\ }\href
  {http://stacks.iop.org/1742-5468/2009/i=02/a=P02035} {\bibfield  {journal}
  {\bibinfo  {journal} {J. Stat. Mech. Theor. Exp.}\ }\textbf {\bibinfo
  {volume} {2009}},\ \bibinfo {pages} {P02035} (\bibinfo {year}
  {2009})}\BibitemShut {NoStop}%
\bibitem [{\citenamefont {Zotos}\ \emph {et~al.}(1997)\citenamefont {Zotos},
  \citenamefont {Naef},\ and\ \citenamefont {Prelov\v{s}ek}}]{Zotos1997}%
  \BibitemOpen
  \bibfield  {author} {\bibinfo {author} {\bibfnamefont {X.}~\bibnamefont
  {Zotos}}, \bibinfo {author} {\bibfnamefont {F.}~\bibnamefont {Naef}}, \ and\
  \bibinfo {author} {\bibfnamefont {P.}~\bibnamefont {Prelov\v{s}ek}},\ }\href
  {\doibase 10.1103/PhysRevB.55.11029} {\bibfield  {journal} {\bibinfo
  {journal} {Phys. Rev. B}\ }\textbf {\bibinfo {volume} {55}},\ \bibinfo
  {pages} {11029} (\bibinfo {year} {1997})}\BibitemShut {NoStop}%
\bibitem [{\citenamefont {Saito}\ and\ \citenamefont
  {Dhar}(2011)}]{Saito2011a}%
  \BibitemOpen
  \bibfield  {author} {\bibinfo {author} {\bibfnamefont {K.}~\bibnamefont
  {Saito}}\ and\ \bibinfo {author} {\bibfnamefont {A.}~\bibnamefont {Dhar}},\
  }\href {\doibase 10.1103/PhysRevE.83.041121} {\bibfield  {journal} {\bibinfo
  {journal} {Phys. Rev. E}\ }\textbf {\bibinfo {volume} {83}},\ \bibinfo
  {pages} {041121} (\bibinfo {year} {2011})}\BibitemShut {NoStop}%
\bibitem [{\citenamefont {Bertini}\ \emph {et~al.}(2015)\citenamefont
  {Bertini}, \citenamefont {{De Sole}}, \citenamefont {Gabrielli},
  \citenamefont {Jona-Lasinio},\ and\ \citenamefont {Landim}}]{Bertini2015}%
  \BibitemOpen
  \bibfield  {author} {\bibinfo {author} {\bibfnamefont {L.}~\bibnamefont
  {Bertini}}, \bibinfo {author} {\bibfnamefont {A.}~\bibnamefont {{De Sole}}},
  \bibinfo {author} {\bibfnamefont {D.}~\bibnamefont {Gabrielli}}, \bibinfo
  {author} {\bibfnamefont {G.}~\bibnamefont {Jona-Lasinio}}, \ and\ \bibinfo
  {author} {\bibfnamefont {C.}~\bibnamefont {Landim}},\ }\href {\doibase
  10.1103/RevModPhys.87.593} {\bibfield  {journal} {\bibinfo  {journal} {Rev.
  Mod. Phys.}\ }\textbf {\bibinfo {volume} {87}},\ \bibinfo {pages} {593}
  (\bibinfo {year} {2015})}\BibitemShut {NoStop}%
\bibitem [{\citenamefont {Bertini}\ \emph {et~al.}(2009)\citenamefont
  {Bertini}, \citenamefont {{De Sole}}, \citenamefont {Gabrielli},
  \citenamefont {Jona-Lasinio},\ and\ \citenamefont {Landim}}]{Bertini2009}%
  \BibitemOpen
  \bibfield  {author} {\bibinfo {author} {\bibfnamefont {L.}~\bibnamefont
  {Bertini}}, \bibinfo {author} {\bibfnamefont {A.}~\bibnamefont {{De Sole}}},
  \bibinfo {author} {\bibfnamefont {D.}~\bibnamefont {Gabrielli}}, \bibinfo
  {author} {\bibfnamefont {G.}~\bibnamefont {Jona-Lasinio}}, \ and\ \bibinfo
  {author} {\bibfnamefont {C.}~\bibnamefont {Landim}},\ }\href {\doibase
  10.1007/s10955-008-9670-4} {\bibfield  {journal} {\bibinfo  {journal} {J.
  Stat. Phys.}\ }\textbf {\bibinfo {volume} {135}},\ \bibinfo {pages} {857}
  (\bibinfo {year} {2009})}\BibitemShut {NoStop}%
\bibitem [{\citenamefont {Bertini}\ \emph {et~al.}(2003)\citenamefont
  {Bertini}, \citenamefont {{De Sole}},\ and\ \citenamefont
  {Gabrielli}}]{Bertini2003}%
  \BibitemOpen
  \bibfield  {author} {\bibinfo {author} {\bibfnamefont {L.}~\bibnamefont
  {Bertini}}, \bibinfo {author} {\bibfnamefont {A.}~\bibnamefont {{De Sole}}},
  \ and\ \bibinfo {author} {\bibfnamefont {D.}~\bibnamefont {Gabrielli}},\
  }\href {\doibase 10.1023/A:1024967818899} {\bibfield  {journal} {\bibinfo
  {journal} {Math. Phys. Anal. Geom.}\ }\textbf {\bibinfo {volume} {6}},\
  \bibinfo {pages} {231} (\bibinfo {year} {2003})}\BibitemShut {NoStop}%
\bibitem [{\citenamefont {Bertini}\ \emph {et~al.}(2007)\citenamefont
  {Bertini}, \citenamefont {{De Sole}}, \citenamefont {Gabrielli},
  \citenamefont {Jona-Lasinio},\ and\ \citenamefont {Landim}}]{Bertini2007}%
  \BibitemOpen
  \bibfield  {author} {\bibinfo {author} {\bibfnamefont {L.}~\bibnamefont
  {Bertini}}, \bibinfo {author} {\bibfnamefont {A.}~\bibnamefont {{De Sole}}},
  \bibinfo {author} {\bibfnamefont {D.}~\bibnamefont {Gabrielli}}, \bibinfo
  {author} {\bibfnamefont {G.}~\bibnamefont {Jona-Lasinio}}, \ and\ \bibinfo
  {author} {\bibfnamefont {C.}~\bibnamefont {Landim}},\ }\href
  {http://stacks.iop.org/1742-5468/2007/i=07/a=P07014} {\bibfield  {journal}
  {\bibinfo  {journal} {J. Stat. Mech. Theor. Exp.}\ }\textbf {\bibinfo
  {volume} {2007}},\ \bibinfo {pages} {P07014} (\bibinfo {year}
  {2007})}\BibitemShut {NoStop}%
\bibitem [{\citenamefont {Aminov}\ \emph {et~al.}(2014)\citenamefont {Aminov},
  \citenamefont {Bunin},\ and\ \citenamefont {Kafri}}]{Aminov2014}%
  \BibitemOpen
  \bibfield  {author} {\bibinfo {author} {\bibfnamefont {A.}~\bibnamefont
  {Aminov}}, \bibinfo {author} {\bibfnamefont {G.}~\bibnamefont {Bunin}}, \
  and\ \bibinfo {author} {\bibfnamefont {Y.}~\bibnamefont {Kafri}},\ }\href
  {\doibase 10.1088/1742-5468/2014/08/P08017} {\bibfield  {journal} {\bibinfo
  {journal} {J. Stat. Mech. Theor. Exp.}\ }\textbf {\bibinfo {volume} {2014}},\
  \bibinfo {pages} {P08017} (\bibinfo {year} {2014})}\BibitemShut {NoStop}%
\bibitem [{\citenamefont {Baek}\ and\ \citenamefont {Kafri}(2015)}]{Baek2015}%
  \BibitemOpen
  \bibfield  {author} {\bibinfo {author} {\bibfnamefont {Y.}~\bibnamefont
  {Baek}}\ and\ \bibinfo {author} {\bibfnamefont {Y.}~\bibnamefont {Kafri}},\
  }\href {http://stacks.iop.org/1742-5468/2015/i=8/a=P08026} {\bibfield
  {journal} {\bibinfo  {journal} {J. Stat. Mech. Theor. Exp.}\ }\textbf
  {\bibinfo {volume} {2015}},\ \bibinfo {pages} {P08026} (\bibinfo {year}
  {2015})}\BibitemShut {NoStop}%
\bibitem [{\citenamefont {Bertini}\ \emph {et~al.}(2013)\citenamefont
  {Bertini}, \citenamefont {Gabrielli}, \citenamefont {Jona-Lasinio},\ and\
  \citenamefont {Landim}}]{Bertini2013}%
  \BibitemOpen
  \bibfield  {author} {\bibinfo {author} {\bibfnamefont {L.}~\bibnamefont
  {Bertini}}, \bibinfo {author} {\bibfnamefont {D.}~\bibnamefont {Gabrielli}},
  \bibinfo {author} {\bibfnamefont {G.}~\bibnamefont {Jona-Lasinio}}, \ and\
  \bibinfo {author} {\bibfnamefont {C.}~\bibnamefont {Landim}},\ }\href
  {\doibase 10.1103/PhysRevLett.110.020601} {\bibfield  {journal} {\bibinfo
  {journal} {Phys. Rev. Lett.}\ }\textbf {\bibinfo {volume} {110}},\ \bibinfo
  {pages} {020601} (\bibinfo {year} {2013})}\BibitemShut {NoStop}%
\bibitem [{\citenamefont {Krapivsky}\ \emph {et~al.}(2012)\citenamefont
  {Krapivsky}, \citenamefont {Meerson},\ and\ \citenamefont
  {Sasorov}}]{Krapivsky2012a}%
  \BibitemOpen
  \bibfield  {author} {\bibinfo {author} {\bibfnamefont {P.~L.}\ \bibnamefont
  {Krapivsky}}, \bibinfo {author} {\bibfnamefont {B.}~\bibnamefont {Meerson}},
  \ and\ \bibinfo {author} {\bibfnamefont {P.~V.}\ \bibnamefont {Sasorov}},\
  }\href {http://stacks.iop.org/1742-5468/2012/i=12/a=P12014} {\bibfield
  {journal} {\bibinfo  {journal} {J. Stat. Mech. Theor. Exp.}\ }\textbf
  {\bibinfo {volume} {2012}},\ \bibinfo {pages} {P12014} (\bibinfo {year}
  {2012})}\BibitemShut {NoStop}%
\bibitem [{\citenamefont {Akkermans}\ \emph {et~al.}(2013)\citenamefont
  {Akkermans}, \citenamefont {Bodineau}, \citenamefont {Derrida},\ and\
  \citenamefont {Shpielberg}}]{Akkermans2013}%
  \BibitemOpen
  \bibfield  {author} {\bibinfo {author} {\bibfnamefont {E.}~\bibnamefont
  {Akkermans}}, \bibinfo {author} {\bibfnamefont {T.}~\bibnamefont {Bodineau}},
  \bibinfo {author} {\bibfnamefont {B.}~\bibnamefont {Derrida}}, \ and\
  \bibinfo {author} {\bibfnamefont {O.}~\bibnamefont {Shpielberg}},\ }\href
  {\doibase 10.1209/0295-5075/103/20001} {\bibfield  {journal} {\bibinfo
  {journal} {Europhys. Lett.}\ }\textbf {\bibinfo {volume} {103}},\ \bibinfo
  {pages} {20001} (\bibinfo {year} {2013})}\BibitemShut {NoStop}%
\bibitem [{\citenamefont {Bodineau}\ \emph {et~al.}(2010)\citenamefont
  {Bodineau}, \citenamefont {Derrida},\ and\ \citenamefont
  {Lebowitz}}]{Bodineau2010a}%
  \BibitemOpen
  \bibfield  {author} {\bibinfo {author} {\bibfnamefont {T.}~\bibnamefont
  {Bodineau}}, \bibinfo {author} {\bibfnamefont {B.}~\bibnamefont {Derrida}}, \
  and\ \bibinfo {author} {\bibfnamefont {J.~L.}\ \bibnamefont {Lebowitz}},\
  }\href {\doibase 10.1007/s10955-010-0012-y} {\bibfield  {journal} {\bibinfo
  {journal} {J. Stat. Phys.}\ }\textbf {\bibinfo {volume} {140}},\ \bibinfo
  {pages} {648} (\bibinfo {year} {2010})}\BibitemShut {NoStop}%
\bibitem [{\citenamefont {Bodineau}\ \emph {et~al.}(2008)\citenamefont
  {Bodineau}, \citenamefont {Derrida},\ and\ \citenamefont
  {Lebowitz}}]{Bodineau2008a}%
  \BibitemOpen
  \bibfield  {author} {\bibinfo {author} {\bibfnamefont {T.}~\bibnamefont
  {Bodineau}}, \bibinfo {author} {\bibfnamefont {B.}~\bibnamefont {Derrida}}, \
  and\ \bibinfo {author} {\bibfnamefont {J.~L.}\ \bibnamefont {Lebowitz}},\
  }\href {\doibase 10.1007/s10955-008-9518-y} {\bibfield  {journal} {\bibinfo
  {journal} {J. Stat. Phys.}\ }\textbf {\bibinfo {volume} {131}},\ \bibinfo
  {pages} {821} (\bibinfo {year} {2008})}\BibitemShut {NoStop}%
\bibitem [{\citenamefont {Bodineau}\ and\ \citenamefont
  {Lagouge}(2010)}]{Bodineau2010}%
  \BibitemOpen
  \bibfield  {author} {\bibinfo {author} {\bibfnamefont {T.}~\bibnamefont
  {Bodineau}}\ and\ \bibinfo {author} {\bibfnamefont {M.}~\bibnamefont
  {Lagouge}},\ }\href {\doibase 10.1007/s10955-010-9934-7} {\bibfield
  {journal} {\bibinfo  {journal} {J. Stat. Phys.}\ }\textbf {\bibinfo {volume}
  {139}},\ \bibinfo {pages} {201} (\bibinfo {year} {2010})}\BibitemShut
  {NoStop}%
\bibitem [{\citenamefont {Agranov}\ \emph {et~al.}(2016)\citenamefont
  {Agranov}, \citenamefont {Meerson},\ and\ \citenamefont
  {Vilenkin}}]{Agranov2016}%
  \BibitemOpen
  \bibfield  {author} {\bibinfo {author} {\bibfnamefont {T.}~\bibnamefont
  {Agranov}}, \bibinfo {author} {\bibfnamefont {B.}~\bibnamefont {Meerson}}, \
  and\ \bibinfo {author} {\bibfnamefont {A.}~\bibnamefont {Vilenkin}},\ }\href
  {\doibase 10.1103/PhysRevE.93.012136} {\bibfield  {journal} {\bibinfo
  {journal} {Phys. Rev. E}\ }\textbf {\bibinfo {volume} {93}},\ \bibinfo
  {pages} {012136} (\bibinfo {year} {2016})}\BibitemShut {NoStop}%
\bibitem [{\citenamefont {Krapivsky}\ \emph {et~al.}(2014)\citenamefont
  {Krapivsky}, \citenamefont {Mallick},\ and\ \citenamefont
  {Sadhu}}]{Krapivsky2014a}%
  \BibitemOpen
  \bibfield  {author} {\bibinfo {author} {\bibfnamefont {P.~L.}\ \bibnamefont
  {Krapivsky}}, \bibinfo {author} {\bibfnamefont {K.}~\bibnamefont {Mallick}},
  \ and\ \bibinfo {author} {\bibfnamefont {T.}~\bibnamefont {Sadhu}},\ }\href
  {\doibase 10.1103/PhysRevLett.113.078101} {\bibfield  {journal} {\bibinfo
  {journal} {Phys. Rev. Lett.}\ }\textbf {\bibinfo {volume} {113}},\ \bibinfo
  {pages} {078101} (\bibinfo {year} {2014})}\BibitemShut {NoStop}%
\bibitem [{\citenamefont {Derrida}\ \emph {et~al.}(2004)\citenamefont
  {Derrida}, \citenamefont {Dou{\c{c}}ot},\ and\ \citenamefont
  {Roche}}]{Derrida2004}%
  \BibitemOpen
  \bibfield  {author} {\bibinfo {author} {\bibfnamefont {B.}~\bibnamefont
  {Derrida}}, \bibinfo {author} {\bibfnamefont {B.}~\bibnamefont
  {Dou{\c{c}}ot}}, \ and\ \bibinfo {author} {\bibfnamefont {P.-E.}\
  \bibnamefont {Roche}},\ }\href
  {http://link.springer.com/10.1023/B:JOSS.0000022379.95508.b2} {\bibfield
  {journal} {\bibinfo  {journal} {J. Stat. Phys.}\ }\textbf {\bibinfo {volume}
  {115}},\ \bibinfo {pages} {717} (\bibinfo {year} {2004})}\BibitemShut
  {NoStop}%
\bibitem [{\citenamefont {Gorissen}\ and\ \citenamefont
  {Vanderzande}(2012)}]{Gorissen2012}%
  \BibitemOpen
  \bibfield  {author} {\bibinfo {author} {\bibfnamefont {M.}~\bibnamefont
  {Gorissen}}\ and\ \bibinfo {author} {\bibfnamefont {C.}~\bibnamefont
  {Vanderzande}},\ }\href {\doibase 10.1103/PhysRevE.86.051114} {\bibfield
  {journal} {\bibinfo  {journal} {Phys. Rev. E}\ }\textbf {\bibinfo {volume}
  {86}},\ \bibinfo {pages} {051114} (\bibinfo {year} {2012})}\BibitemShut
  {NoStop}%
\bibitem [{\citenamefont {Appert-Rolland}\ \emph {et~al.}(2008)\citenamefont
  {Appert-Rolland}, \citenamefont {Derrida}, \citenamefont {Lecomte},\ and\
  \citenamefont {van Wijland}}]{Appert-Rolland2008}%
  \BibitemOpen
  \bibfield  {author} {\bibinfo {author} {\bibfnamefont {C.}~\bibnamefont
  {Appert-Rolland}}, \bibinfo {author} {\bibfnamefont {B.}~\bibnamefont
  {Derrida}}, \bibinfo {author} {\bibfnamefont {V.}~\bibnamefont {Lecomte}}, \
  and\ \bibinfo {author} {\bibfnamefont {F.}~\bibnamefont {van Wijland}},\
  }\href {\doibase 10.1103/PhysRevE.78.021122} {\bibfield  {journal} {\bibinfo
  {journal} {Phys. Rev. E}\ }\textbf {\bibinfo {volume} {78}},\ \bibinfo
  {pages} {021122} (\bibinfo {year} {2008})}\BibitemShut {NoStop}%
\bibitem [{\citenamefont {Tiz\'{o}n-Escamilla}\ \emph
  {et~al.}(2016{\natexlab{a}})\citenamefont {Tiz\'{o}n-Escamilla},
  \citenamefont {Hurtado},\ and\ \citenamefont {Garrido}}]{Tizon16a}%
  \BibitemOpen
  \bibfield  {author} {\bibinfo {author} {\bibfnamefont {N.}~\bibnamefont
  {Tiz\'{o}n-Escamilla}}, \bibinfo {author} {\bibfnamefont {P.~I.}\
  \bibnamefont {Hurtado}}, \ and\ \bibinfo {author} {\bibfnamefont {P.~L.}\
  \bibnamefont {Garrido}},\ }\href@noop {} {\  (\bibinfo {year}
  {2016}{\natexlab{a}})},\ \Eprint {http://arxiv.org/abs/1611.02500}
  {arXiv:1611.02500 [cond-mat]} \BibitemShut {NoStop}%
\bibitem [{\citenamefont {Tiz\'{o}n-Escamilla}\ \emph
  {et~al.}(2016{\natexlab{b}})\citenamefont {Tiz\'{o}n-Escamilla},
  \citenamefont {Hurtado},\ and\ \citenamefont {Garrido}}]{Tizon16b}%
  \BibitemOpen
  \bibfield  {author} {\bibinfo {author} {\bibfnamefont {N.}~\bibnamefont
  {Tiz\'{o}n-Escamilla}}, \bibinfo {author} {\bibfnamefont {P.~I.}\
  \bibnamefont {Hurtado}}, \ and\ \bibinfo {author} {\bibfnamefont {P.~L.}\
  \bibnamefont {Garrido}},\ }\href@noop {} {\  (\bibinfo {year}
  {2016}{\natexlab{b}})},\ \Eprint {http://arxiv.org/abs/1606.07507}
  {arXiv:1606.07507 [cond-mat]} \BibitemShut {NoStop}%
\bibitem [{\citenamefont {Akkermans}\ and\ \citenamefont
  {Montambaux}(2007)}]{Akkermans2007}%
  \BibitemOpen
  \bibfield  {author} {\bibinfo {author} {\bibfnamefont {E.}~\bibnamefont
  {Akkermans}}\ and\ \bibinfo {author} {\bibfnamefont {G.}~\bibnamefont
  {Montambaux}},\ }\href {\doibase 10.1017/CBO9780511618833} {\emph {\bibinfo
  {title} {Mesoscopic Physics of Electrons and Photons:}}}\ (\bibinfo
  {publisher} {Cambridge University Press},\ \bibinfo {year}
  {2007})\BibitemShut {NoStop}%
\bibitem [{\citenamefont {Kamenev}(2011)}]{Kamenev2011}%
  \BibitemOpen
  \bibfield  {author} {\bibinfo {author} {\bibfnamefont {A.}~\bibnamefont
  {Kamenev}},\ }\href
  {http://www.cambridge.org/pl/academic/subjects/physics/condensed-matter-physics-nanoscience-and-mesoscopic-physics/field-theory-non-equilibrium-systems?format=HB}
  {\emph {\bibinfo {title} {Field Theory of Non-Equilibrium Systems}}}\
  (\bibinfo  {publisher} {Cambridge University Press},\ \bibinfo {year}
  {2011})\BibitemShut {NoStop}%
\bibitem [{\citenamefont {Jordan}\ \emph {et~al.}(2004)\citenamefont {Jordan},
  \citenamefont {Sukhorukov},\ and\ \citenamefont {Pilgram}}]{Jordan2004}%
  \BibitemOpen
  \bibfield  {author} {\bibinfo {author} {\bibfnamefont {A.~N.}\ \bibnamefont
  {Jordan}}, \bibinfo {author} {\bibfnamefont {E.~V.}\ \bibnamefont
  {Sukhorukov}}, \ and\ \bibinfo {author} {\bibfnamefont {S.}~\bibnamefont
  {Pilgram}},\ }\href {\doibase http://dx.doi.org/10.1063/1.1803927} {\bibfield
   {journal} {\bibinfo  {journal} {J. Math. Phys.}\ }\textbf {\bibinfo {volume}
  {45}},\ \bibinfo {pages} {4386} (\bibinfo {year} {2004})}\BibitemShut
  {NoStop}%
\bibitem [{\citenamefont {Kambly}\ \emph {et~al.}(2011)\citenamefont {Kambly},
  \citenamefont {Flindt},\ and\ \citenamefont {B\"uttiker}}]{Kambly2011}%
  \BibitemOpen
  \bibfield  {author} {\bibinfo {author} {\bibfnamefont {D.}~\bibnamefont
  {Kambly}}, \bibinfo {author} {\bibfnamefont {C.}~\bibnamefont {Flindt}}, \
  and\ \bibinfo {author} {\bibfnamefont {M.}~\bibnamefont {B\"uttiker}},\
  }\href {\doibase 10.1103/PhysRevB.83.075432} {\bibfield  {journal} {\bibinfo
  {journal} {Phys. Rev. B}\ }\textbf {\bibinfo {volume} {83}},\ \bibinfo
  {pages} {075432} (\bibinfo {year} {2011})}\BibitemShut {NoStop}%
\bibitem [{\citenamefont {Bertini}\ \emph {et~al.}(2006)\citenamefont
  {Bertini}, \citenamefont {{De Sole}}, \citenamefont {Gabrielli},
  \citenamefont {Jona-Lasinio},\ and\ \citenamefont {Landim}}]{Bertini2006}%
  \BibitemOpen
  \bibfield  {author} {\bibinfo {author} {\bibfnamefont {L.}~\bibnamefont
  {Bertini}}, \bibinfo {author} {\bibfnamefont {A.}~\bibnamefont {{De Sole}}},
  \bibinfo {author} {\bibfnamefont {D.}~\bibnamefont {Gabrielli}}, \bibinfo
  {author} {\bibfnamefont {G.}~\bibnamefont {Jona-Lasinio}}, \ and\ \bibinfo
  {author} {\bibfnamefont {C.}~\bibnamefont {Landim}},\ }\href {\doibase
  10.1007/s10955-006-9056-4} {\bibfield  {journal} {\bibinfo  {journal} {J.
  Stat. Phys.}\ }\textbf {\bibinfo {volume} {123}},\ \bibinfo {pages} {237}
  (\bibinfo {year} {2006})}\BibitemShut {NoStop}%
\bibitem [{\citenamefont {Bodineau}\ and\ \citenamefont
  {Derrida}(2004)}]{Bodineau2004}%
  \BibitemOpen
  \bibfield  {author} {\bibinfo {author} {\bibfnamefont {T.}~\bibnamefont
  {Bodineau}}\ and\ \bibinfo {author} {\bibfnamefont {B.}~\bibnamefont
  {Derrida}},\ }\href {\doibase 10.1103/PhysRevLett.92.180601} {\bibfield
  {journal} {\bibinfo  {journal} {Phys. Rev. Lett.}\ }\textbf {\bibinfo
  {volume} {92}},\ \bibinfo {pages} {180601} (\bibinfo {year}
  {2004})}\BibitemShut {NoStop}%
\bibitem [{\citenamefont {Imparato}\ \emph {et~al.}(2009)\citenamefont
  {Imparato}, \citenamefont {Lecomte},\ and\ \citenamefont {van
  Wijland}}]{Imparato2009}%
  \BibitemOpen
  \bibfield  {author} {\bibinfo {author} {\bibfnamefont {A.}~\bibnamefont
  {Imparato}}, \bibinfo {author} {\bibfnamefont {V.}~\bibnamefont {Lecomte}}, \
  and\ \bibinfo {author} {\bibfnamefont {F.}~\bibnamefont {van Wijland}},\
  }\href {\doibase 10.1103/PhysRevE.80.011131} {\bibfield  {journal} {\bibinfo
  {journal} {Phys. Rev. E}\ }\textbf {\bibinfo {volume} {80}},\ \bibinfo
  {pages} {011131} (\bibinfo {year} {2009})}\BibitemShut {NoStop}%
\bibitem [{\citenamefont {Hurtado}\ and\ \citenamefont
  {Garrido}(2009)}]{Hurtado2009}%
  \BibitemOpen
  \bibfield  {author} {\bibinfo {author} {\bibfnamefont {P.}~\bibnamefont
  {Hurtado}}\ and\ \bibinfo {author} {\bibfnamefont {P.}~\bibnamefont
  {Garrido}},\ }\href {\doibase 10.1103/PhysRevLett.102.250601} {\bibfield
  {journal} {\bibinfo  {journal} {Phys. Rev. Lett.}\ }\textbf {\bibinfo
  {volume} {102}},\ \bibinfo {pages} {250601} (\bibinfo {year}
  {2009})}\BibitemShut {NoStop}%
\bibitem [{\citenamefont {Hurtado}\ and\ \citenamefont
  {Garrido}(2010)}]{Hurtado2010}%
  \BibitemOpen
  \bibfield  {author} {\bibinfo {author} {\bibfnamefont {P.~I.}\ \bibnamefont
  {Hurtado}}\ and\ \bibinfo {author} {\bibfnamefont {P.~L.}\ \bibnamefont
  {Garrido}},\ }\href {\doibase 10.1103/PhysRevE.81.041102} {\bibfield
  {journal} {\bibinfo  {journal} {Phys. Rev. E}\ }\textbf {\bibinfo {volume}
  {81}},\ \bibinfo {pages} {041102} (\bibinfo {year} {2010})}\BibitemShut
  {NoStop}%
\bibitem [{\citenamefont {Baek}\ \emph {et~al.}(2016)\citenamefont {Baek},
  \citenamefont {Kafri},\ and\ \citenamefont {Lecomte}}]{Baek2016b}%
  \BibitemOpen
  \bibfield  {author} {\bibinfo {author} {\bibfnamefont {Y.}~\bibnamefont
  {Baek}}, \bibinfo {author} {\bibfnamefont {Y.}~\bibnamefont {Kafri}}, \ and\
  \bibinfo {author} {\bibfnamefont {V.}~\bibnamefont {Lecomte}},\ }\href@noop
  {} {\  (\bibinfo {year} {2016})},\ \Eprint {http://arxiv.org/abs/1609.06732}
  {arXiv:1609.06732 [cond-mat]} \BibitemShut {NoStop}%
\bibitem [{Note1()}]{Note1}%
  \BibitemOpen
  \bibinfo {note} {However, for periodic boundary conditions, many models were
  shown to break the AP assumption explicitly, see~\cite {Appert-Rolland2008}
  and~\cite {Zarfaty2015a,Bodineau2005,Hurtado2011a}.}\BibitemShut {Stop}%
\bibitem [{\citenamefont {Bertini}\ \emph {et~al.}(2005)\citenamefont
  {Bertini}, \citenamefont {De~Sole}, \citenamefont {Gabrielli}, \citenamefont
  {Jona-Lasinio},\ and\ \citenamefont {Landim}}]{Bertini2005}%
  \BibitemOpen
  \bibfield  {author} {\bibinfo {author} {\bibfnamefont {L.}~\bibnamefont
  {Bertini}}, \bibinfo {author} {\bibfnamefont {A.}~\bibnamefont {De~Sole}},
  \bibinfo {author} {\bibfnamefont {D.}~\bibnamefont {Gabrielli}}, \bibinfo
  {author} {\bibfnamefont {G.}~\bibnamefont {Jona-Lasinio}}, \ and\ \bibinfo
  {author} {\bibfnamefont {C.}~\bibnamefont {Landim}},\ }\href {\doibase
  10.1103/PhysRevLett.94.030601} {\bibfield  {journal} {\bibinfo  {journal}
  {Phys. Rev. Lett.}\ }\textbf {\bibinfo {volume} {94}},\ \bibinfo {pages}
  {030601} (\bibinfo {year} {2005})}\BibitemShut {NoStop}%
\bibitem [{\citenamefont {Shpielberg}\ and\ \citenamefont
  {Akkermans}(2016)}]{Shpielberg2016}%
  \BibitemOpen
  \bibfield  {author} {\bibinfo {author} {\bibfnamefont {O.}~\bibnamefont
  {Shpielberg}}\ and\ \bibinfo {author} {\bibfnamefont {E.}~\bibnamefont
  {Akkermans}},\ }\href {\doibase 10.1103/PhysRevLett.116.240603} {\bibfield
  {journal} {\bibinfo  {journal} {Phys. Rev. Lett.}\ }\textbf {\bibinfo
  {volume} {116}},\ \bibinfo {pages} {240603} (\bibinfo {year}
  {2016})}\BibitemShut {NoStop}%
\bibitem [{\citenamefont {Landau}\ \emph {et~al.}(1980)\citenamefont {Landau},
  \citenamefont {Lifshitz},\ and\ \citenamefont
  {Pitaevskii}}]{LandauLifshiftz1980StatisticalPt1}%
  \BibitemOpen
  \bibfield  {author} {\bibinfo {author} {\bibfnamefont {L.~D.}\ \bibnamefont
  {Landau}}, \bibinfo {author} {\bibfnamefont {E.~M.}\ \bibnamefont
  {Lifshitz}}, \ and\ \bibinfo {author} {\bibfnamefont {L.~P.}\ \bibnamefont
  {Pitaevskii}},\ }\href {https://books.google.co.il/books?id=sZUfAQAAMAAJ}
  {\emph {\bibinfo {title} {Statistical Physics, Part 1}}},\ \bibinfo {series}
  {Course of Theoretical Physics}, Vol.~\bibinfo {volume} {5}\ (\bibinfo
  {publisher} {Pergamon Press},\ \bibinfo {address} {Oxford},\ \bibinfo {year}
  {1980})\BibitemShut {NoStop}%
\bibitem [{\citenamefont {Zarfaty}\ and\ \citenamefont
  {Meerson}(2016)}]{Zarfaty2015a}%
  \BibitemOpen
  \bibfield  {author} {\bibinfo {author} {\bibfnamefont {L.}~\bibnamefont
  {Zarfaty}}\ and\ \bibinfo {author} {\bibfnamefont {B.}~\bibnamefont
  {Meerson}},\ }\href {http://stacks.iop.org/1742-5468/2016/i=3/a=033304}
  {\bibfield  {journal} {\bibinfo  {journal} {J. Stat. Mech. Theor. Exp.}\
  }\textbf {\bibinfo {volume} {2016}},\ \bibinfo {pages} {033304} (\bibinfo
  {year} {2016})}\BibitemShut {NoStop}%
\bibitem [{\citenamefont {Bodineau}\ and\ \citenamefont
  {Derrida}(2005)}]{Bodineau2005}%
  \BibitemOpen
  \bibfield  {author} {\bibinfo {author} {\bibfnamefont {T.}~\bibnamefont
  {Bodineau}}\ and\ \bibinfo {author} {\bibfnamefont {B.}~\bibnamefont
  {Derrida}},\ }\href {\doibase 10.1103/PhysRevE.72.066110} {\bibfield
  {journal} {\bibinfo  {journal} {Phys. Rev. E}\ }\textbf {\bibinfo {volume}
  {72}},\ \bibinfo {pages} {066110} (\bibinfo {year} {2005})}\BibitemShut
  {NoStop}%
\bibitem [{\citenamefont {Kipnis}\ \emph {et~al.}(1982)\citenamefont {Kipnis},
  \citenamefont {Marchioro},\ and\ \citenamefont {Presutti}}]{Kipnis1982}%
  \BibitemOpen
  \bibfield  {author} {\bibinfo {author} {\bibfnamefont {C.}~\bibnamefont
  {Kipnis}}, \bibinfo {author} {\bibfnamefont {C.}~\bibnamefont {Marchioro}}, \
  and\ \bibinfo {author} {\bibfnamefont {E.}~\bibnamefont {Presutti}},\ }\href
  {\doibase 10.1007/BF01011740} {\bibfield  {journal} {\bibinfo  {journal} {J.
  Stat. Phys.}\ }\textbf {\bibinfo {volume} {27}},\ \bibinfo {pages} {65}
  (\bibinfo {year} {1982})}\BibitemShut {NoStop}%
\bibitem [{\citenamefont {Bodineau}()}]{BodineauPrivate}%
  \BibitemOpen
  \bibfield  {author} {\bibinfo {author} {\bibfnamefont {T.}~\bibnamefont
  {Bodineau}},\ }\href@noop {} {}\bibinfo {note} {{private communication (see
  the Supplementary Materials of~\cite{Shpielberg2016})}}\BibitemShut {NoStop}%
\bibitem [{\citenamefont {Derrida}(2007)}]{Derrida2007}%
  \BibitemOpen
  \bibfield  {author} {\bibinfo {author} {\bibfnamefont {B.}~\bibnamefont
  {Derrida}},\ }\href {http://stacks.iop.org/1742-5468/2007/i=07/a=P07023}
  {\bibfield  {journal} {\bibinfo  {journal} {J. Stat. Mech. Theor. Exp.}\
  }\textbf {\bibinfo {volume} {2007}},\ \bibinfo {pages} {P07023} (\bibinfo
  {year} {2007})}\BibitemShut {NoStop}%
\bibitem [{\citenamefont {Tailleur}\ \emph {et~al.}(2007)\citenamefont
  {Tailleur}, \citenamefont {Kurchan},\ and\ \citenamefont
  {Lecomte}}]{Tailleur07}%
  \BibitemOpen
  \bibfield  {author} {\bibinfo {author} {\bibfnamefont {J.}~\bibnamefont
  {Tailleur}}, \bibinfo {author} {\bibfnamefont {J.}~\bibnamefont {Kurchan}}, \
  and\ \bibinfo {author} {\bibfnamefont {V.}~\bibnamefont {Lecomte}},\ }\href
  {\doibase 10.1103/PhysRevLett.89.030601???} {\bibfield  {journal} {\bibinfo
  {journal} {Phys. Rev. Lett.}\ }\textbf {\bibinfo {volume} {99}},\ \bibinfo
  {pages} {150602} (\bibinfo {year} {2007})}\BibitemShut {NoStop}%
\bibitem [{Note2()}]{Note2}%
  \BibitemOpen
  \bibinfo {note} {Here we assume the number of particles in the system is
  bounded.}\BibitemShut {Stop}%
\bibitem [{Note3()}]{Note3}%
  \BibitemOpen
  \bibinfo {note} {We will also consider $p$ to be time-independent, although
  it may not be necessarily so.}\BibitemShut {Stop}%
\bibitem [{Note4()}]{Note4}%
  \BibitemOpen
  \bibinfo {note} {This scheme can only probe small fluctuations about the AP
  solution. To our knowledge, there are no sufficient and necessary conditions
  which accommodate large fluctuations.}\BibitemShut {Stop}%
\bibitem [{Note5()}]{Note5}%
  \BibitemOpen
  \bibinfo {note} {Where it is assumed there is no accumulation of particles in
  the system. See~\cite {Hirschberg2015} for an example where this assumption
  does not hold.}\BibitemShut {Stop}%
\bibitem [{\citenamefont {Spohn}(1991)}]{Spohn1992Part2Chap2}%
  \BibitemOpen
  \bibfield  {author} {\bibinfo {author} {\bibfnamefont {H.}~\bibnamefont
  {Spohn}},\ }\enquote {\bibinfo {title} {Equilibrium fluctuations},}\ in\
  \href {\doibase 10.1007/978-3-642-84371-6} {\emph {\bibinfo {booktitle}
  {{Large Scale Dynamics of Interacting Particles}}}},\ \bibinfo {series and
  number} {Texts and Monographs in Physics}\ (\bibinfo  {publisher} {Springer
  Berlin Heidelberg},\ \bibinfo {address} {Berlin, Heidelberg},\ \bibinfo
  {year} {1991})\ \bibinfo {type} {{Part II},}\ \bibinfo {chapter} {Chap.\ 2},
  pp.\ \bibinfo {pages} {175--211}\BibitemShut {NoStop}%
\bibitem [{\citenamefont {Garnier}\ and\ \citenamefont
  {Ciliberto}(2005)}]{Garnier2005}%
  \BibitemOpen
  \bibfield  {author} {\bibinfo {author} {\bibfnamefont {N.}~\bibnamefont
  {Garnier}}\ and\ \bibinfo {author} {\bibfnamefont {S.}~\bibnamefont
  {Ciliberto}},\ }\href {\doibase 10.1103/PhysRevE.71.060101} {\bibfield
  {journal} {\bibinfo  {journal} {Phys. Rev. E}\ }\textbf {\bibinfo {volume}
  {71}},\ \bibinfo {pages} {060101} (\bibinfo {year} {2005})}\BibitemShut
  {NoStop}%
\bibitem [{\citenamefont {Hurtado}\ and\ \citenamefont
  {Garrido}(2011)}]{Hurtado2011a}%
  \BibitemOpen
  \bibfield  {author} {\bibinfo {author} {\bibfnamefont {P.~I.}\ \bibnamefont
  {Hurtado}}\ and\ \bibinfo {author} {\bibfnamefont {P.~L.}\ \bibnamefont
  {Garrido}},\ }\href {\doibase 10.1103/PhysRevLett.107.180601} {\bibfield
  {journal} {\bibinfo  {journal} {Phys. Rev. Lett.}\ }\textbf {\bibinfo
  {volume} {107}},\ \bibinfo {pages} {180601} (\bibinfo {year}
  {2011})}\BibitemShut {NoStop}%
\bibitem [{\citenamefont {Hirschberg}\ \emph {et~al.}(2015)\citenamefont
  {Hirschberg}, \citenamefont {Mukamel},\ and\ \citenamefont
  {Sch\"utz}}]{Hirschberg2015}%
  \BibitemOpen
  \bibfield  {author} {\bibinfo {author} {\bibfnamefont {O.}~\bibnamefont
  {Hirschberg}}, \bibinfo {author} {\bibfnamefont {D.}~\bibnamefont {Mukamel}},
  \ and\ \bibinfo {author} {\bibfnamefont {G.~M.}\ \bibnamefont {Sch\"utz}},\
  }\href {http://stacks.iop.org/1742-5468/2015/i=11/a=P11023} {\bibfield
  {journal} {\bibinfo  {journal} {J. Stat. Mech. Theor. Exp.}\ }\textbf
  {\bibinfo {volume} {2015}},\ \bibinfo {pages} {P11023} (\bibinfo {year}
  {2015})}\BibitemShut {NoStop}%
\end{thebibliography}

% ======================================================================================== %

\end{document}